\documentclass[11pt,showpacs,notitlepage]{revtex4-1} % for 2 columns style with line numbers
\usepackage{graphicx}
\usepackage{bm}
\usepackage{amssymb}
\usepackage{amsmath}
\usepackage{fullpage}

\begin{document}

\title{Reduction of a metapopulation genetic model to an effective one island model}

%\shorttitle{Reduction of a metapopulation genetic model to an effective model}

\author{C\'esar Parra-Rojas}
\affiliation{Frankfurt Institute for Advanced Studies, 60438 Frankfurt am Main, Germany}
\author{Alan~J. McKane}
\affiliation{Theoretical Physics Division, School of Physics and Astronomy, The University of Manchester, Manchester M13 9PL, UK}
%
%
%\date{}
%\shortauthor{C\'{e}sar Parra-Rojas \etal}

%\institute{                    
%  \inst{1} Frankfurt Institute for Advanced Studies, 60438 Frankfurt am Main, Germany\\
%  \inst{2} Theoretical Physics Division, School of Physics and Astronomy, 
%The University of Manchester, Manchester M13 9PL, UK
%}

%\abstract{}
\begin{abstract}
We explore a model of metapopulation genetics which is based on a more ecologically motivated approach than is frequently used in population genetics. The size of the population is regulated by competition between individuals, rather than by artificially imposing a fixed population size. The increased complexity of the model is managed by employing techniques often used in the physical sciences, namely exploiting time-scale separation to eliminate fast variables and then constructing an effective model from the slow modes. We analyse this effective model and show that the predictions for the probability of fixation of the alleles and the mean time to fixation agree well with those found from numerical simulations of the original model.
\end{abstract}

\maketitle

\section{Introduction}
The subject of population genetics holds a particular fascination for statistical physicists because of the many analogies it has with various models in non-equilibrium statistical mechanics~\cite{deViadar2011,black2012}. Much of the formalism used by physicists in the study of non-equilibrium systems derives from viewing these as stochastic processes, and is directly applicable to the investigation of models of population genetics~\cite{ewens1969,ewens2004}. The concepts that are frequently of interest there, such as the probability that a particular allele fixes, and the mean time to fixation, are also the focus of attention in many physical systems out of equilibrium~\cite{gardiner2009,risken1989}.

As the genetic models have become increasingly complicated, incorporating spatial structure, sexual reproduction or several gene loci, the methods of solution previously employed are no longer efficacious. The purpose of this article is to describe a systematic method for reducing the full models to effective models, which still provide good predictions for quantities relating to the fixation of alleles, but which are simple enough to be analysed mathematically. This method has previously been applied to several models of population genetics; here we apply it to a model not previously considered. In this way we hope that the article has the dual function of serving as a concise review of the approach, but also providing some original results.

The specific model we will discuss will have a spatial aspect: several subpopulations in distinct regions, with individuals able to migrate to one region from another. It will therefore have many parameters: birth, death and competition rates which differ between alleles and between regions; the regions, in turn, vary in size (in the sense that they can sustain different numbers of individuals), and the migration rates between them are also variable. We are therefore confronted with the difficulty of analysing a rather complex model, as discussed above. This is managed by making two approximations, which we will show give excellent agreement with results found by simulating the original model. 

The first is the standard diffusion approximation~\cite{crow2009}, which in the language of statistical physics consists of moving from the microscopic description in terms of individuals to a mesoscopic description in terms the fraction of the population in the various regions that is of one type or the other. The second approximation is the neglect of degrees of freedom that decay rapidly on time scales that are of interest to us. This approximation also has a long history, and is known variously as adiabatic elimination~\cite{Haken1983}, fast variable elimination~\cite{vanKampen1985}, centre manifold (CM) theory~\cite{wiggins2003}, among others. In the present application it will turn out that all degrees of freedom but one, decay away relatively quickly, leaving an effective theory which is sufficiently simple to be analytically tractable.

The modelling procedure that we will adopt will include the effects of migration, selection and genetic drift, but the processes of birth and death will be taken to be distinct, unlike the conventional approach in population genetics where birth and death are coupled in order to keep the population size fixed~\cite{fisher1930,wright1931,moran1958}. Instead, a competition between the individuals in the system will be introduced that will have the effect of keeping the population fixed on average, but with ever present fluctuations about this average. In this way the basic elements of the model will more closely resemble an ecological model with the processes of birth, death and competition, but where the different species are identified by the fact that they carry different alleles. We will only examine the case of a single gene in haploid individuals that can only have two variants; we will refer to the alleles as type $1$ and type $2$. The method can be extended to diploid and multiallelic individuals, but here we prefer to focus on the effects of spatial structure, selection due to varying birth, death and competition rates between the species, and genetic drift due to stochastic effects resulting from the finite number of individuals present in the system.

We seek to make the model as generic as possible, and so we will construct it at the fundamental level of individuals undergoing the processes of birth, death, competition and migration. The simplest choices for these processes lead to a Lotka-Volterra competition model~\cite{pielou1977}, and since the model will be stochastic, we will refer to it as a stochastic Lotka-Volterra competition (SLVC) model. The spatial structure will be introduced by asking that the population is divided into $\mathcal{D}$ subpopulations in distinct regions. In population genetics these might be referred to as demes or islands; here we will use the terminology of islands, following the practice in ecology. Similarly we will refer to the population as a metapopulation~\cite{hanski1999}, since it will have the structure of a network where the nodes are islands, with different sizes and with varying link strength (level of migration) between them. 

\section{Model}
As we have stressed above, we believe it is important to begin at the level of discrete individuals and the interactions between them. As also mentioned, in common with most authors, we make the diffusion approximation~\cite{crow2009} in order to make progress in analysing the model. Within this approximation the variables are the number density of individuals of type $\alpha$ on island $i$, denoted by $x^{(\alpha)}_i$. The parameters of the model are both local (the birth and death rates of these individuals, respectively $b^{(\alpha)}_i$ and $d^{(\alpha)}_i$, and the competition between types $\alpha$ and $\beta$ on island $i$ denoted by $c^{(\alpha \beta)}_i$) and non-local (the rate $\mu_{ij}$ at which an individual from island $j$ will migrate to island $i$). The specification of the model and the application of the diffision approximation is by now standard~\cite{crow2009,vanKampen2007}, and is discussed in detail for this particular model in Sec.~\ref{sec:setup_SM} of the supplementary material (SM). Our interest here is in the second approximation discussed in the introduction, which can be made after this first approximation has been carried out, and therefore our starting point will be the stochastic differential equation which is the outcome of the analysis described in Sec.~\ref{sec:setup_SM} of the SM.

To simplify the form of the stochastic differential equation it is useful to introduce an index $I$ that runs from $1$ to $2\mathcal{D}$, so that $I=i$ if the allele labelled is $1$ and if the island being considered is $i$, and $I=\mathcal{D}+i$ if the allele labelled is $2$ and if the island being considered is $i$. The state of the system is denoted by the vector $\bm{x} = (x^{(1)}_1,x^{(2)}_1,\ldots,x^{(\mathcal{D})}_1,x^{(1)}_2,x^{(2)}_2,\ldots,x^{(\mathcal{D})}_2)$. As discussed in Sec.~\ref{sec:setup_SM} of the SM, the model also contains a set of $\mathcal{D}$ parameters, $V_i$, which denote the potential capacity of island $i$, both in terms of environmental factors required to sustain a population and the size of the island. Within the diffusion approximation we set $V_i = \beta_i V$, where $\beta_i$ is a number of order one that characterises the capacity of each island compared to the others, and where $V$ is the typical carrying capacity of an island, which is the central parameter which controls the diffusion approximation. After these definitions, we may now write the stochastic differential equation (defined in the sense of It\={o}~\cite{gardiner2009}) derived in the SM in the form 
\begin{equation}
\frac{\mathrm{d}x_I}{\mathrm{d}\tau} = A_I(\bm{x}) + \frac{1}{\sqrt{V}} \eta_I(\tau),
\label{SDE_full}
\end{equation}
where $\tau = t/V$ is a rescaled time and $\eta_I(\tau)$ is a Gaussian white noise with zero mean and with a correlator 
\begin{equation}
\left\langle \eta_I(\tau) \eta_J(\tau') \right\rangle = B_{I J}(\bm{x}) \delta\left( \tau - \tau' \right).
\label{correlator}
\end{equation}
The functions $A_I(\bm{x})$ and $B_{I J}(\bm{x})$ which specify the model, are derived in Sec.~\ref{sec:setup_SM} of the SM, beginning from the microscopic description given by Eqs.~(\ref{trans_rates})-(\ref{stoichiometric}). They are given by
\begin{eqnarray}
A^{(\alpha)}_i(\bm{x}) &=& \frac{1}{\beta_i} \left\{ \left( b^{(\alpha)}_i - d^{(\alpha)}_i \right)x^{(\alpha)}_i - \sum^{2}_{\beta = 1} c^{(\alpha \beta)}_i x^{(\alpha)}_i x^{(\beta)}_i+ \mathcal{M}^{(\alpha,-)}_i \right\}, \ \ \ \ i=1,\ldots,\mathcal{D}, \ \ \ \alpha=1,2,\nonumber \\
\label{full_A}
\end{eqnarray}
and
\begin{eqnarray}
B^{(\alpha \alpha)}_{ii}(\bm{x}) &=& \frac{1}{\beta^2_i} \left\{ \left( b^{(\alpha)}_i + d^{(\alpha)}_i \right)x^{(\alpha)}_i + \sum^{2}_{\beta = 1} c^{(\alpha \beta)}_i x^{(\alpha)}_i x^{(\beta)}_i + \mathcal{M}^{(\alpha,+)}_i \right\}, \ \ \ \ i=1,\ldots,\mathcal{D}, \ \ \ \alpha=1,2,\nonumber \\
\label{part_of_B}
\end{eqnarray}
where the nonlocal contributions due to migration, $\mathcal{M}^{(\alpha,\pm)}_i$, are given by
\begin{equation}
\mathcal{M}^{(\alpha,\pm)}_i = \sum_{j \neq i} \left[ \mu_{ij} x^{(\alpha)}_j \pm \mu_{ji} x^{(\alpha)}_i \right].
\label{mig_plus_minus}
\end{equation}
In addition,
\begin{equation}
B^{(\alpha \alpha)}_{ij}(\bm{x}) = - \frac{1}{\beta_i \beta_j}\,\left[ \mu_{ij} x^{(\alpha)}_j + \mu_{ji} x^{(\alpha)}_i \right], \ \ \left(i \neq j\right),
\label{rest_of_B}
\end{equation}
and $B^{(12)}_{ij}=B^{(21)}_{ij}=0$, for all $i,j$.

While the transition rates which define the model at the level of individuals (given by Eq.~(\ref{trans_rates})) are rather transparent, and can be written down from the model description, the forms of the equivalent mesoscopic quantities $A_I(\bm{x})$ and $B_{IJ}(\bm{x})$ given above are rather less clear. The $A_I(\bm{x})$, from which the deterministic dynamics follow, has some familiar elements, namely the first two terms in the curly brackets which are the usual Lotka-Volterra interaction terms. So although analytic progress is helped by making the diffusion approximation, the fact that the functions given by Eqs.~(\ref{full_A}) and (\ref{part_of_B}) are still very complex, means that further approximations are required. We will now show that the elimination of fast modes is an approximation which can be justified biologically, and yields a simplified model which retains the power to make accurate predictions for quantities such as probabilities of fixation and mean fixation times.

\section{Identification of slow modes}
In this second approximation, the mesoscopic model with $2\mathcal{D}$ degrees of freedom may be reduced to one with only a single degree of freedom. This reduced model can essentially be thought of as one with no spatial structure, but defined by a set of \textit{effective} parameters, which encapsulate those of the full model. Later we will compare the result of calculations from the reduced model to numerical simulations of the original.

%%%%%%%%%%%%%%%%%%%%%%%%%%%%%%%%%%%%%%%%%%%%%%%%%%%%%%%%%%%%%%%%%%%%%%%%%%%%%%%
%%%%%%%%%%%%%%%%%%%%%%%%%%%%%%%%%%%%%%%%%%%%%%%%%%%%%%%%%%%%%%%%%%%%%%%%%%%%%%%

\begin{figure}[h]
\centering
\includegraphics[width=0.48\columnwidth]{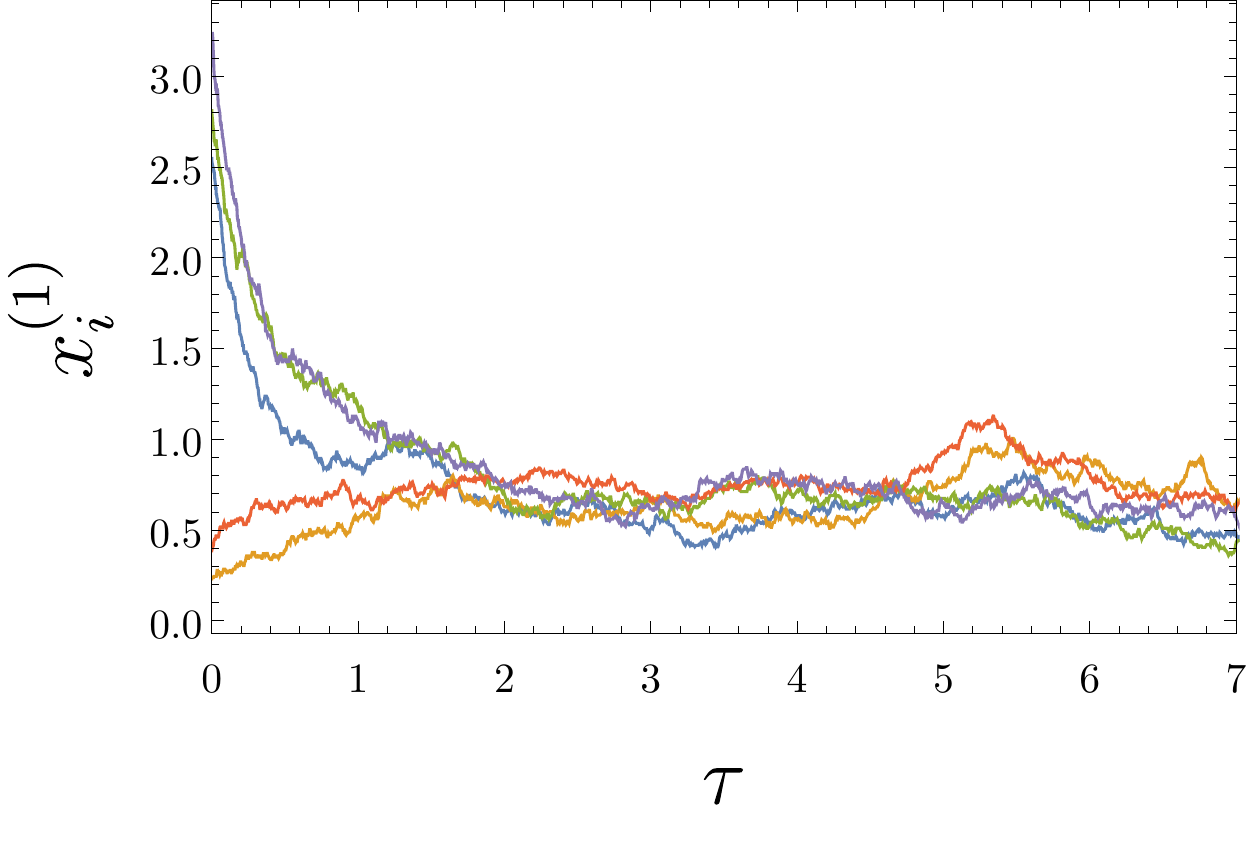}\hfill
\includegraphics[width=0.48\columnwidth]{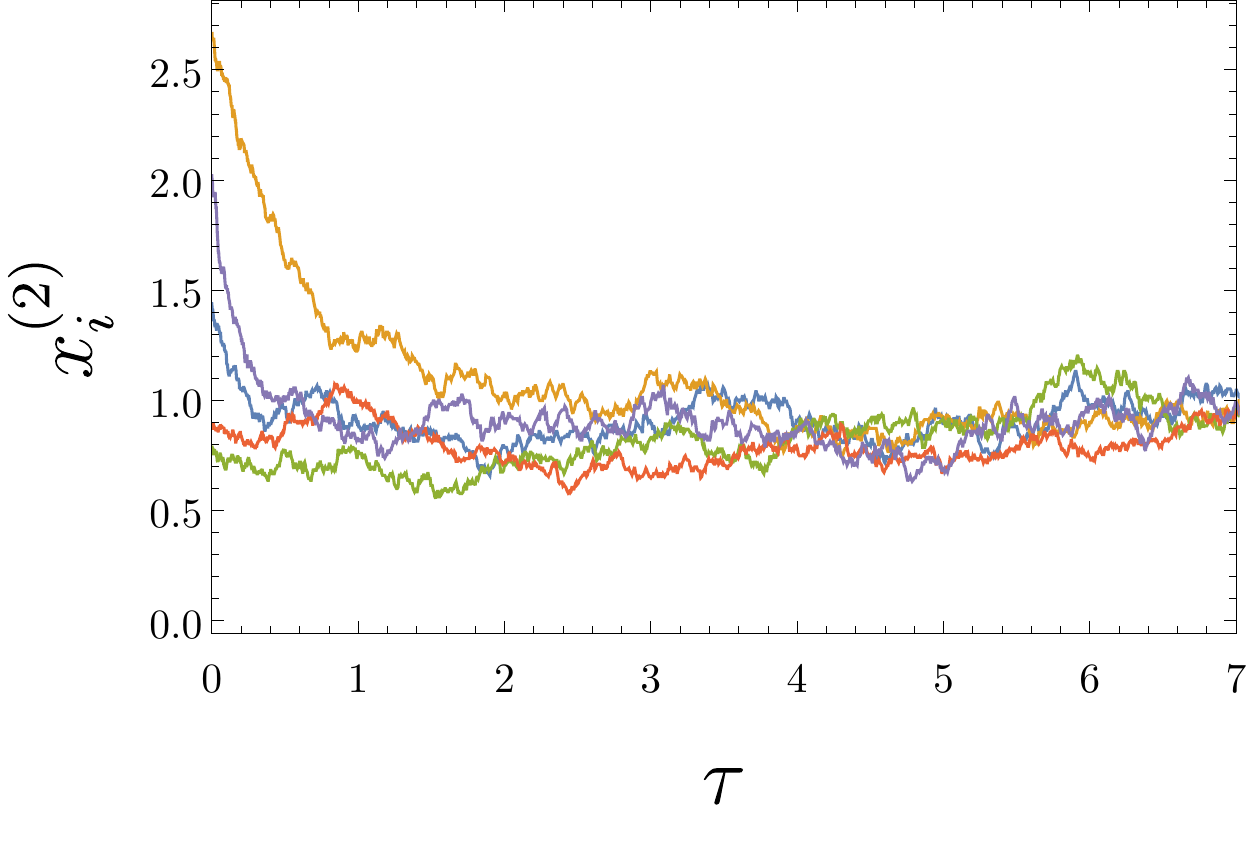}
\caption{Collapse in the fractions of individuals of type $1$ (left) and 2 (right) towards an island-independent trajectory in the neutral case. The number of islands is $\mathcal{D}=5$, and each line corresponds to a single stochastic trajectory of $x^{(\alpha)}_i$, with $i=1,\ldots,5$. Parameters: $V=150$, $\kappa=1.5$.}
\label{fig:collapse}
\end{figure}

%%%%%%%%%%%%%%%%%%%%%%%%%%%%%%%%%%%%%%%%%%%%%%%%%%%%%%%%%%%%%%%%%%%%%%%%%%%%%%%
%%%%%%%%%%%%%%%%%%%%%%%%%%%%%%%%%%%%%%%%%%%%%%%%%%%%%%%%%%%%%%%%%%%%%%%%%%%%%%%

The method is based on the observation that the dynamics of the full model consists of two stages. The first consists of a relatively rapid decay from the initial state to the vicinity of a CM (if selection is absent) or a slow subspace (SS) (if selection is present). It then enters the second stage where it wanders stochastically on or near the CM (and also weakly deterministically on a SS if weak selection is present) until fixation of one or other of the alleles; this is shown in Fig.~\ref{fig:collapse} for a neutral system with $\mathcal{D}=5$ islands. This is the heart of the time-scale separation: the rate of migration, which controls the collapse onto the SS, is much greater than the rate of genetic drift, which eventually leads to global fixation. Time-scale separation arguments have also been used on similar models elsewhere~\cite{lombardo_2014,lombardo_2015}. In the dynamics of the first stage, stochastic effects play very little role; there is what is in essence a deterministic collapse onto the CM (or SS). We will therefore study this first stage of the process deterministically, beginning with the case of no selection, where a true CM exists.

\subsection{Neutral model}
In SLVC models, selection is introduced through the parameters $b^{(\alpha)}_i, d^{(\alpha)}_i$, and $c^{(\alpha \beta)}_i$, which if made to vary with $\alpha$ and $\beta$, give a selective advantage to those individuals carrying either allele $\alpha$ or allele $\beta$. Therefore to have no selection we set $b^{(\alpha)}_i = b^{(0)}_i$, $d^{(\alpha)}_i = d^{(0)}_i$, and $c^{(\alpha \beta)}_i = c^{(0)}_i$ for all $\alpha, \beta=1,2$. Substituting this into the deterministic equation $\mathrm{d}x_I/\mathrm{d}\tau = A_I(\bm{x})$, obtained by taking the $V \to \infty$ limit of Eq.~(\ref{SDE_full}), yields
\begin{eqnarray}
\frac{\mathrm{d}x^{(\alpha)}_i}{\mathrm{d}\tau} &=& \frac{1}{\beta_i} \left\{ \left( b^{(0)}_i - d^{(0)}_i \right)x^{(\alpha)}_i - c^{(0)}_i x^{(\alpha)}_i \sum^{2}_{\beta = 1} x^{(\beta)}_i + \mathcal{M}^{(\alpha,-)}_i \right\}, \ \ \ \ i=1,\ldots,\mathcal{D}, \ \ \ \alpha=1,2.
\label{neutral_deterministic}
\end{eqnarray}
To achieve the maximum reduction, we are searching for a low-dimensional CM. In this case we can find one which is one-dimensional, by seeking fixed points of Eq.~(\ref{neutral_deterministic}) that are independent of $i$, that is, solutions of
\begin{equation}
x^{(\alpha)} \left[ \left( b^{(0)}_i + q_i - d^{(0)}_i \right) - c^{(0)}_i \sum^{2}_{\beta = 1} x^{(\beta)} \right] = 0, \ \ \ \alpha=1,2,
\label{neutral_FPs}
\end{equation}
where
\begin{equation}
q_i \equiv  \sum_{j \neq i} \left[ \mu_{ij} - \mu_{ji} \right].
\label{q_i_defn}
\end{equation}
The only solution of Eq.~(\ref{neutral_FPs}), apart from the trivial solution $x^{(1)}=x^{(2)}=0$, is 
\begin{equation}
x^{(1)} + x^{(2)} = \frac{\left( b^{(0)}_i + q_i - d^{(0)}_i \right)}{c^{(0)}_i},
\label{CM_x}
\end{equation}
which, for consistency, requires that $( b^{(0)}_i + q_i - d^{(0)}_i) = \kappa c^{(0)}_i$ for all $i$, where $\kappa$ is a constant. This condition should perhaps not be surprising, since we are reducing the model from one with $2\mathcal{D}$ degrees of freedom to one with only one degree of freedom ($x^{(1)}$, with $x^{(2)}$ determined from Eq.~(\ref{CM_x})). Therefore each island has in some sense to be neutral in order to obtain a neutral one-island model. Later, when we introduce selection, we will be able to move away from this assumption.

\begin{figure}
\centering
\includegraphics[width=0.45\columnwidth]{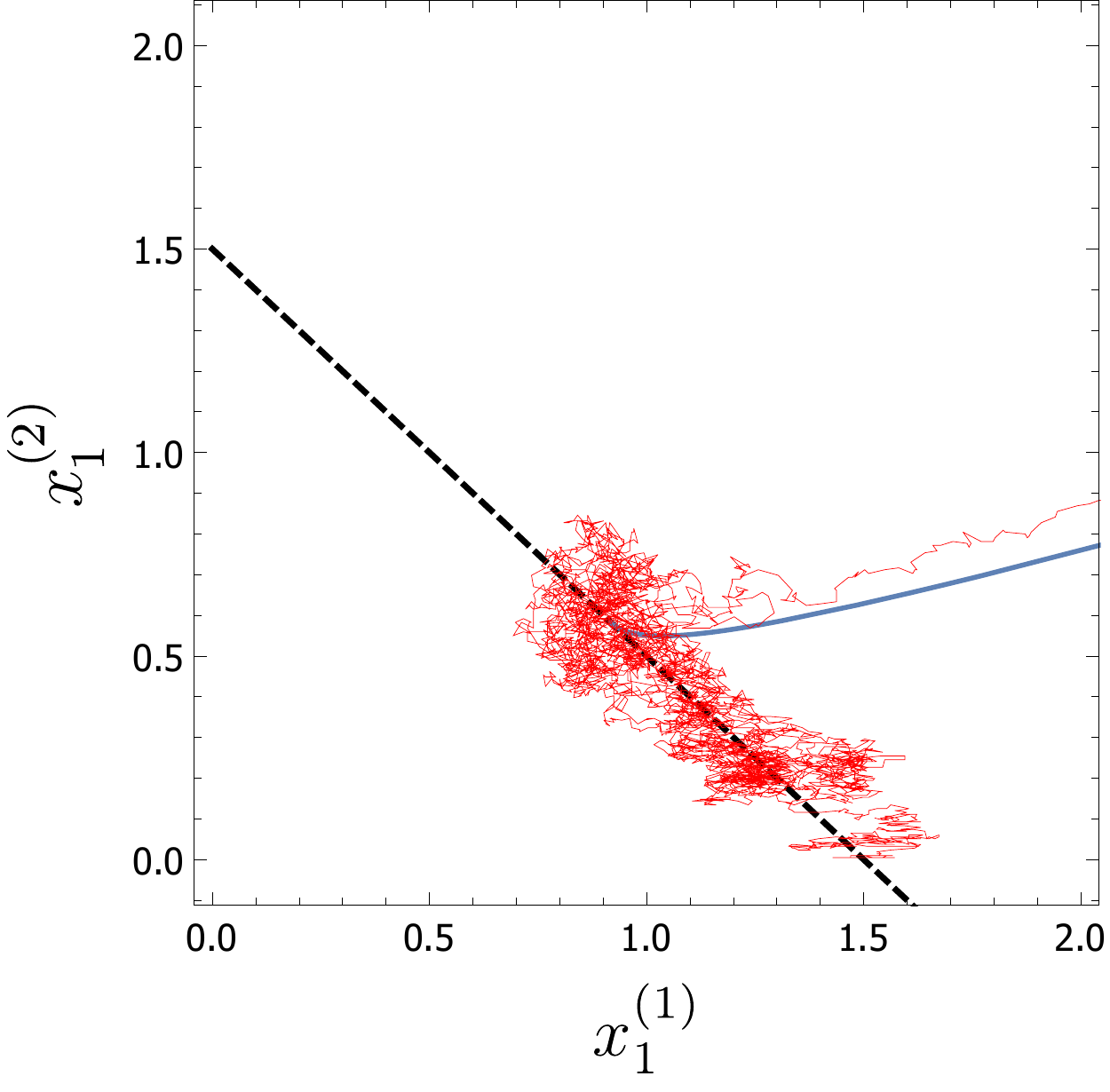}
\caption{A neutral system with two islands. Phase diagram for individuals of types $1$ and $2$ on island 1. Blue line: deterministic trajectory; red line: one stochastic trajectory; black, dashed line: CM given by Eq.~\eqref{CM_x}. Parameters: $V=300$, $\kappa =1.5$.}
\label{fig:phase_neutral}
\end{figure}

Equation (\ref{CM_x}) defines the one-dimensional CM, which we show for a two-island system in the phase diagram of Fig.~\ref{fig:phase_neutral}---Fig.~\ref{fig:phase_neutral_SM} (see SM) further shows that the solution is island-independent. Before proceeding any further, we scale the original variables of the system, in order to make the analysis more transparent. To do this, we define variables 
\begin{equation}
y^{(\alpha)}_i = \frac{c^{(0)}_i x^{(\alpha)}_i}{\left( b^{(0)}_i + q_i - d^{(0)}_i \right)} = \kappa^{-1} x^{(\alpha)}_i,
\label{y_variables_defn}
\end{equation}
with $i=1,\ldots,\mathcal{D}$ and $\alpha=1,2$. Then repeating the analysis of this section, but in the $y^{(\alpha)}_i$ variables, rather than in the $x^{(\alpha)}_i$, we find a CM where $y^{(\alpha)}_i = y^{(\alpha)}$ for all $i$ and $\alpha=1,2$, with
\begin{equation}
y^{(1)} + y^{(2)} = 1.
\label{CM_y}
\end{equation}
We will choose the CM to be parameterised by $y^{(1)}$ which we will denote by $z$, the only variable of the reduced system. Then $y^{(2)} = 1 - z$.

The more complete analysis carried out in Sec.~\ref{sec:slow_modes_SM} of the SM, involves finding the eigenvalues and eigenvectors of the Jacobian on the CM. The decay rates of the modes associated with the various eigenvectors are proportional to the inverse of the corresponding eigenvalues. In the SM the $2\mathcal{D} - 1$ `fast' modes are identified; the single slow mode---which is actually static when there is no selection, since it has eigenvalue zero---corresponds to the CM. For the purposes of the general overview presented here, the fast modes simply take the system from its initial condition (IC) to the CM, the initial point of contact being referred to as the initial condition on the CM (CMIC).  

As discussed earlier in this section, we assume that in this first part of the dynamics---the decay from the initial condition, $\bm{y}^{\rm IC}$, to the CM---the deterministic dynamics completely dominates the stochastic dynamics. In effect, this means that it is assumed that the stochastic system still reaches the CM at the point $z^{\rm CMIC}$ found from the deterministic neutral dynamics, and that this can be used as an initial condition for the second stage of the dynamics, which takes place entirely on the CM. This assumption will be examined in the numerical simulations which are discussed later and in the SM.

%%%%%%%%%%%%%%%%%%%%%%%%%%%%%%%%%%%%%%%%%%%%%%%%%%%%%%%%%%%%%%%%%%%%%%%%%%%%%%%
%%%%%%%%%%%%%%%%%%%%%%%%%%%%%%%%%%%%%%%%%%%%%%%%%%%%%%%%%%%%%%%%%%%%%%%%%%%%%%%

%%%%%%%%%%%%%%%%%%%%%%%%%%%%%%%%%%%%%%%%%%%%%%%%%%%%%%%%%%%%%%%%%%%%%%%%%%%%%%%
%%%%%%%%%%%%%%%%%%%%%%%%%%%%%%%%%%%%%%%%%%%%%%%%%%%%%%%%%%%%%%%%%%%%%%%%%%%%%%%

\subsection{Model with selection}
To go on to analyse the non-neutral case we write the birth, death and competition parameters as follows:
\begin{eqnarray}
& & b^{(\alpha)}_i = b^{(0)}_i \left( 1 + \epsilon \hat{b}^{(\alpha)}_{i} \right);
\ \ d^{(\alpha)}_i = d^{(0)}_i \left( 1 + \epsilon \hat{d}^{(\alpha)}_{i} \right);\ \  c^{(\alpha \beta)}_{i} = c^{(0)}_i \left( 1 + \epsilon \hat{c}^{(\alpha \beta)}_i \right).
\label{parameters_non_neut}
\end{eqnarray}
Here $\epsilon$ is the selection strength. As described in Sec.~\ref{sec:selection_SM} of the SM, we assume that $\epsilon$ and $V^{-1}$ are of the same order, and therefore keep order $\epsilon$ terms in $A_{I}(\bm{y})$, but only order one terms in $B_{IJ}(\bm{y})$. The noise correlator will then correspond to the one obtained from the neutral theory (see Sec.~\ref{sec:neutral_SM} of the SM). 

In order to find $A_{I}(\bm{y})$ to first order in $\epsilon$, we write the coordinates on the SS as
\begin{equation}
y^{(1)}_i = z + \epsilon Y^{(1)}_i + \mathcal{O}\left( \epsilon^2 \right), \ \ \ 
y^{(2)}_i = (1-z) + \epsilon Y^{(2)}_i + \mathcal{O}\left( \epsilon^2 \right),
\label{SS_coords}
\end{equation}
where $Y^{(1)}_i$ and $Y^{(2)}_i$ are to be determined. Substituting these coordinates into the expressions for $A^{(1)}_{i}(\bm{y})$ and $A^{(2)}_{i}(\bm{y})$ (see Eq.~(\ref{full_A})), but restricted to the SS, together with some further analysis, gives Eq.~\eqref{summary_form} for the equation of the SS. 

\section{Construction of the reduced model}
So far we have identified the one-dimensional subspace that the system collapses onto (the SS) and have identified the variable which moves the system along this subspace ($z$). The subspace itself was found by starting from Eq.~(\ref{SS_coords}) and asking that $A_I(\bm{y})$ only had components along the subspace. We can also ask that the noise only acts along the SS; technically this is best achieved through the construction of a projection operator which in effect projects the stochastic differential equation (\ref{SDE_full}) onto a one-dimensional stochastic differential equation consisting of an effective deterministic function $\bar{A}(z)$, with the noise having an effective correlator $\bar{B}(z)$. 

The details of how this projection is carried out are given in the SM where it is shown (see Sec.~\ref{sec:reduction_SM}) that we arrive at the following form for the stochastic differential equation describing the stochastic dynamics after the fast-mode elimination:
\begin{equation}
\frac{\mathrm{d}z}{\mathrm{d}\tau} = \bar{A}(z) + \frac{1}{\sqrt{V}} \zeta(\tau),
\label{SDE_reduced}
\end{equation}
where $\zeta(\tau)$ is a Gaussian noise with zero mean and correlator
\begin{equation}
\left\langle \zeta(\tau) \zeta(\tau') \right\rangle = \bar{B}(z) \delta\left( \tau - \tau' \right)\,.
\label{correlator_reduced}
\end{equation}
Here 
\begin{equation}
\bar{A}(z) = \epsilon z \left( 1 - z \right)\left( a_1 + a_2 z \right) + \mathcal{O}\left( \epsilon^2 \right)\,,
\label{Abar_very_simple}
\end{equation}
where
\begin{eqnarray}
a_1 = \sum^{\mathcal{D}}_{i=1} \frac{u^{\{1\} }_i}{\beta_i}\,\left\{ \left[ \left( b^{(0)}_i \hat{b}^{(1)}_i - d^{(0)}_i \hat{d}^{(1)}_i \right)% \right. \right. & & \nonumber \\ 
- \left( b^{(0)}_i \hat{b}^{(2)}_i - d^{(0)}_i \hat{d}^{(2)}_i \right) \right] + \kappa c^{(0)}_i \left[ \hat{c}^{(2 2)}_i - \hat{c}^{(1 2)}_i \right] \right\} & &   
\label{a_one}
\end{eqnarray}
and
\begin{equation}
a_2 = - \sum^{\mathcal{D}}_{i=1} \frac{\kappa u^{\{1\} }_i c^{(0)}_i \Gamma_i }{\beta_i}\,,
\label{a_two}
\end{equation}
and where we have defined $\Gamma_i \equiv \hat{c}^{(11)}_i - \hat{c}^{(12)}_i - \hat{c}^{(21)}_i + \hat{c}^{(22)}_i$. In addition, $\bm{u}^{\{1\} }$ is the eigenvector of the $\mathcal{D} \times \mathcal{D}$ matrix with off-diagonal elements $\mu_{ij}/\beta_i$ and diagonal elements $- \sum_{j \neq i} \mu_{ij}/\beta_i$, having eigenvalue zero.

In the same way, the reduced noise correlator is found to be (see Eq.~\eqref{form_of_Bbar})
\begin{equation}
\bar{B}(z) = 2bz\left( 1 - z \right),
\label{Bbar_very_simple}
\end{equation}
where
\begin{equation}
b = \kappa^{-1} \sum^{\mathcal{D}}_{i=1} \frac{\left[ u^{\{ 1\} }_i \right]^{2}}{\beta^2_i} b^{(0)}_i\,.
\label{bee}
\end{equation}
We see that the forms for $\bar{A}(z)$ and $\bar{B}(z)$ are similar to those that we might expect from a model with only one degree of freedom, but with the parameters of the model ($a_1,a_2$ and $b$) encapsulating some of the structure of the original $2\mathcal{D}$-degrees-of-freedom model. The reduced stochastic differential equation (\ref{SDE_reduced}), together with the correlation function in Eq.~(\ref{correlator_reduced}) and Eqs.~(\ref{Abar_very_simple}) and (\ref{Bbar_very_simple}), completely describe the stochastic dynamics of the reduced system.

It is straightforward to check that the results obtained above agree with an earlier analysis carried out for a single island, i.e.~$\mathcal{D}=1$~\cite{constable2015a}. In the single-island reduction, a further simplication was made, which while not necessary, does simplify the analysis. This consisted in asking that the SS passes through the two points $\bm{y}=(1,0)$ and $\bm{y}=(0,1)$~\cite{constable2015a}. The analogue in the present case is the requirement that when $z=1$, $y^{(1)}_i=1$ and $y^{(2)}_i=0$, for all $i$. Similarly that when $z=0$, $y^{(1)}_i=0$ and $y^{(2)}_i=1$, for all $i$. If these conditions are not imposed, there is a stochastic drift along the SS until either of the axes is reached and fixation of one of the types is achieved. The imposition of the conditions reduces the number of parameters of the model and ensures that fixation occurs at $z=0$ and $z=1$. In Sec.~\ref{sec:selection_SM} of the SM we show that these conditions imply that 
\begin{eqnarray}
\left( b^{(0)}_i \hat{b}^{(1)}_i - d^{(0)}_i \hat{d}^{(1)}_i \right) &=& \kappa c^{(0)}_i \hat{c}^{(1 1)}_i, \nonumber \\
\left( b^{(0)}_i \hat{b}^{(2)}_i - d^{(0)}_i \hat{d}^{(2)}_i \right) &=& \kappa c^{(0)}_i \hat{c}^{(2 2)}_i,
\label{conds_end_points}
\end{eqnarray}
where $i=1,\ldots,\mathcal{D}$. Substitution of the conditions in Eq.~(\ref{conds_end_points}) into Eq.~(\ref{a_one}), leads to a form for Eq.~(\ref{Abar_very_simple}), at order $\epsilon$, which is given by
\begin{eqnarray}
\bar{A}(z) &=& \epsilon z \left( 1 - z \right) \sum^{\mathcal{D}}_{i=1} \frac{\kappa c^{(0)}_i u^{\{1\} }_i}{\beta_i}\,\left[ \phi^{(1)}_i - \Gamma_i z \right]\,,
\label{Abar_end_cond}
\end{eqnarray}
where
\begin{equation}
\phi^{(1)}_i \equiv  \hat{c}^{(1 1)}_i - \hat{c}^{(1 2)}_i\,.
\label{phi_1_defn}
\end{equation}
This shows that all dependence on the birth and death parameters has been eliminated; the result for $\bar{A}(z)$ only depends on the competition parameters. 

In the same way as was done in the general case, effective parameters, which contain information about the full model, can be introduced:
\begin{eqnarray}
\Gamma_{\rm eff} &\equiv& \sum^{\mathcal{D}}_{i=1} \frac{\kappa c^{(0)}_i u^{\{1\} }_i}{\beta_i}\,\Gamma_i, \nonumber \\
\phi^{(1)}_{\rm eff} &\equiv& \sum^{\mathcal{D}}_{i=1} \frac{\kappa c^{(0)}_i u^{\{1\} }_i}{\beta_i}\,\phi^{(1)}_i\,.
\label{effective_paras}
\end{eqnarray}
This then yields
\begin{equation}
\bar{A}(z) = \epsilon z \left( 1 - z \right)\,\left[ \phi^{(1)}_{\rm eff} - \Gamma_{\rm eff} z + \mathcal{O}\left( \epsilon \right) \right],
\label{Abar_like_Ref_Oned}
\end{equation}
which has the same form as in the one-island case~\cite{constable2015a}, but now with effective parameters. It should be stressed that the simplification leading to Eq.~(\ref{conds_end_points}) was simply made as a special case which leads to a simpler end result, which can be useful in checking the efficacy of the method; the more general form given by Eqs.~(\ref{Abar_very_simple})-(\ref{a_two}) should and can be used in general. 

Figure~\ref{fig:phase_select} shows a phase diagram for a system with $\mathcal{D}=2$ islands and selection. The rather strong level of selection allows us to clearly appreciate the fact that a CM no longer exists, and the system collapses towards a curved SS instead; on the latter, both deterministic and stochastic dynamics take place. In the next section we will use the reduced model to make predictions, and test these through numerical simulation of the original model.

\section{Analysis of the reduced model}
The purpose of this section is twofold. Firstly, to note that the one-degree-of-freedom model given in the previous section can be analysed mathematically, and to compare the predictions of this reduced model to simulations of the full model. Secondly, to use these results to investigate the quality of the approximations made to obtain the reduced model.

%%%%%%%%%%%%%%%%%%%%%%%%%%%%%%%%%%%%%%%%%%%%%%%%%%%%%%%%%%%%%%%%%%%%%%%%%%%%%%%
%%%%%%%%%%%%%%%%%%%%%%%%%%%%%%%%%%%%%%%%%%%%%%%%%%%%%%%%%%%%%%%%%%%%%%%%%%%%%%%

\begin{figure}
\centering
\includegraphics[width=0.45\columnwidth]{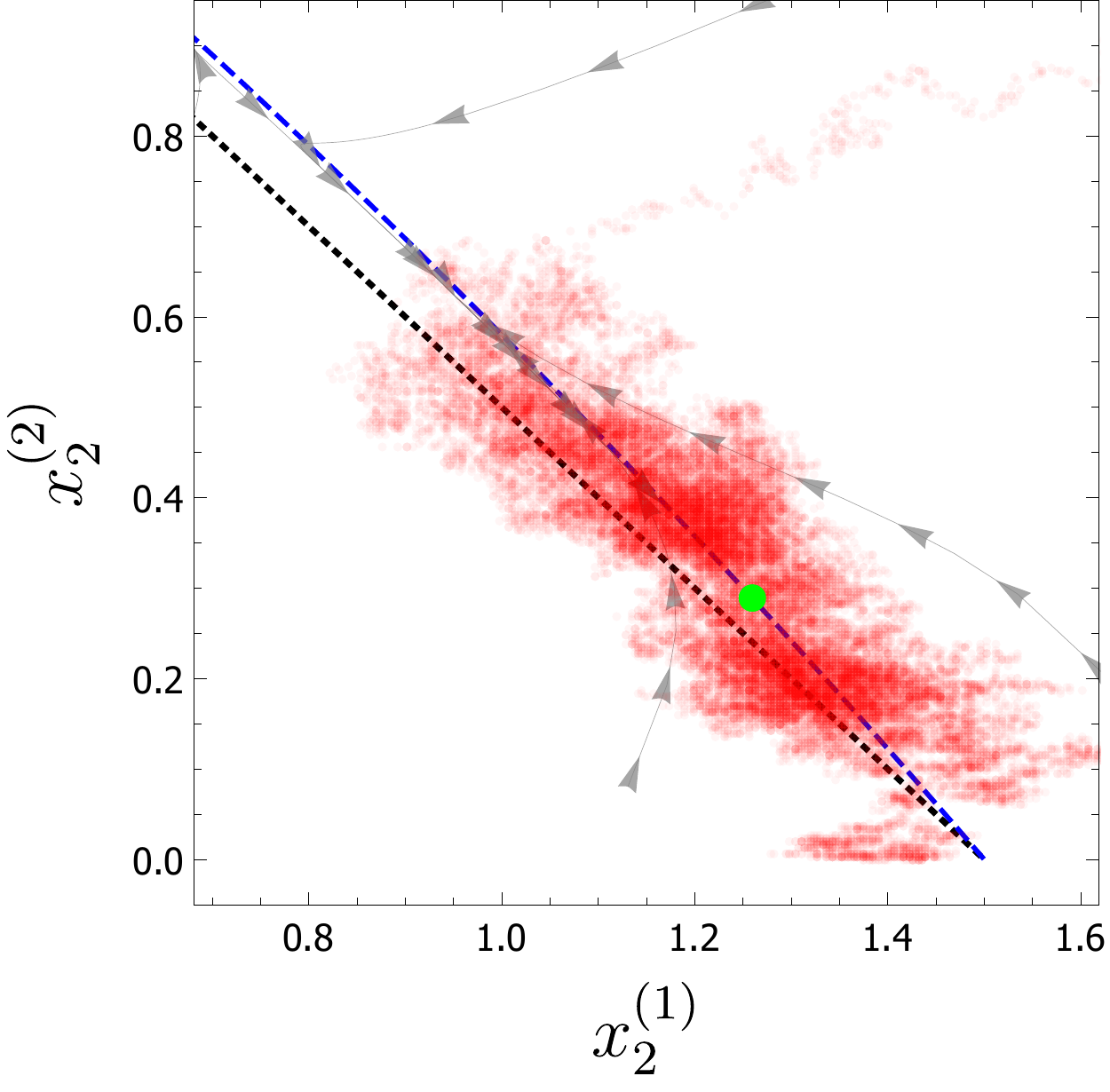}
\caption{A system with two islands and selection. Phase diagram for individuals of type $1$ and $2$ on island 2. Grey lines: deterministic trajectories for different initial conditions; red dots: one stochastic trajectory; black, dotted line: CM from the neutral theory; blue, dashed line: slow subspace; green dot: stable fixed point of the reduced system. Parameters: $V=500$, $\kappa =1.5$, $\epsilon=0.3$, $\phi_{\rm eff}^{(1)}\approx 0.4$, $\phi_{\rm eff}^{(2)}\approx 0.08$, $\Gamma_{\rm eff}\approx 0.48$, $z^*\approx 0.83$.}
\label{fig:phase_select}
\end{figure}

%%%%%%%%%%%%%%%%%%%%%%%%%%%%%%%%%%%%%%%%%%%%%%%%%%%%%%%%%%%%%%%%%%%%%%%%%%%%%%%
%%%%%%%%%%%%%%%%%%%%%%%%%%%%%%%%%%%%%%%%%%%%%%%%%%%%%%%%%%%%%%%%%%%%%%%%%%%%%%%

Although the form of the reduced model closely resembles those of one-dimensional stochastic models in population genetics~\cite{crow2009}, there is one significant difference. This is that $\bar{A}(z)$ is in general cubic in the variable $z$, rather than having a simple quadratic form such as $sz(1-z)$, where $s$ is a selection coefficient. This difference implies that there is a possibility of an `internal' fixed point---one away from the boundaries at $z=0$ and $z=1$. One might naively expect that the presence of a stable fixed point would lead to a longer mean time to fixation and an unstable fixed point to a shorter mean time to fixation.

%%%%%%%%%%%%%%%%%%%%%%%%%%%%%%%%%%%%%%%%%%%%%%%%%%%%%%%%%%%%%%%%%%%%%%%%%%%%%%%
%%%%%%%%%%%%%%%%%%%%%%%%%%%%%%%%%%%%%%%%%%%%%%%%%%%%%%%%%%%%%%%%%%%%%%%%%%%%%%%

\begin{figure}
\centering
\includegraphics[width=0.48\columnwidth]{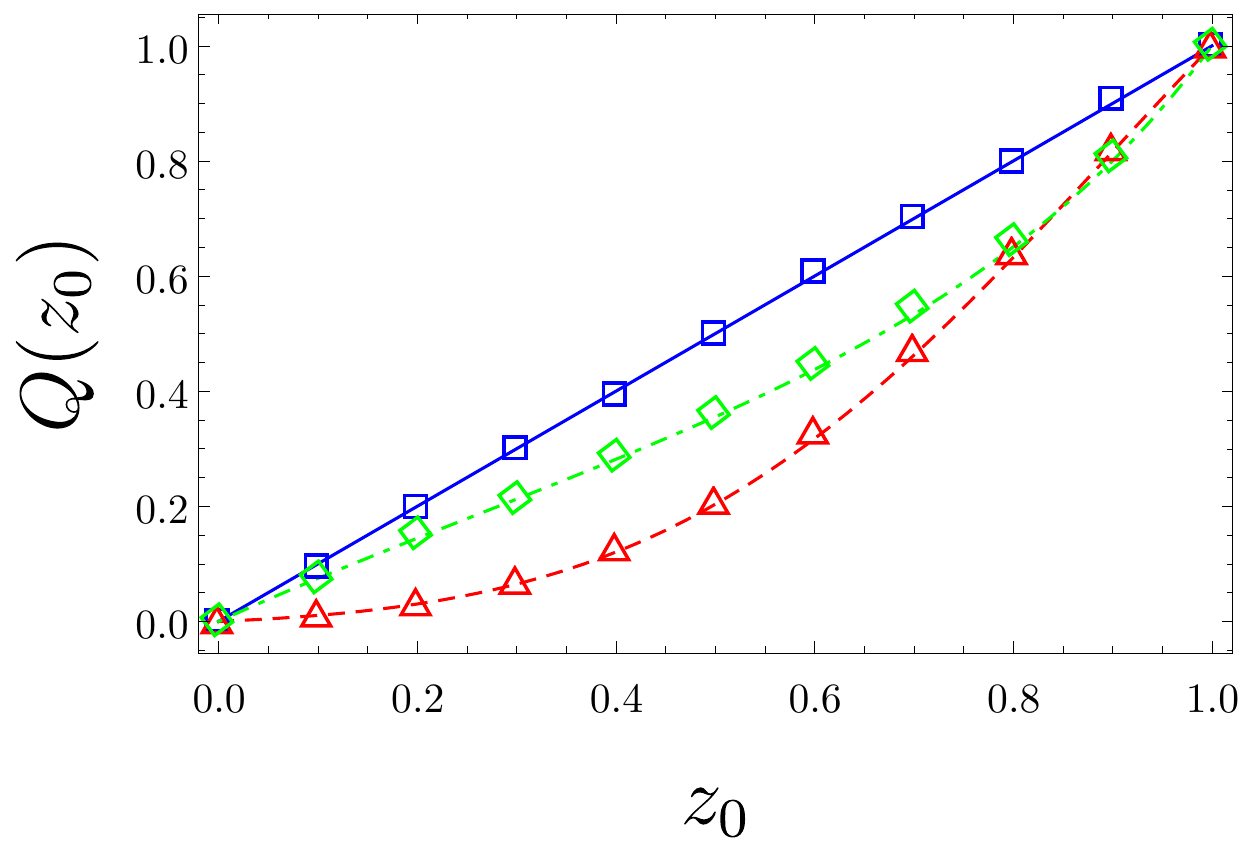}\hfill
\includegraphics[width=0.48\columnwidth]{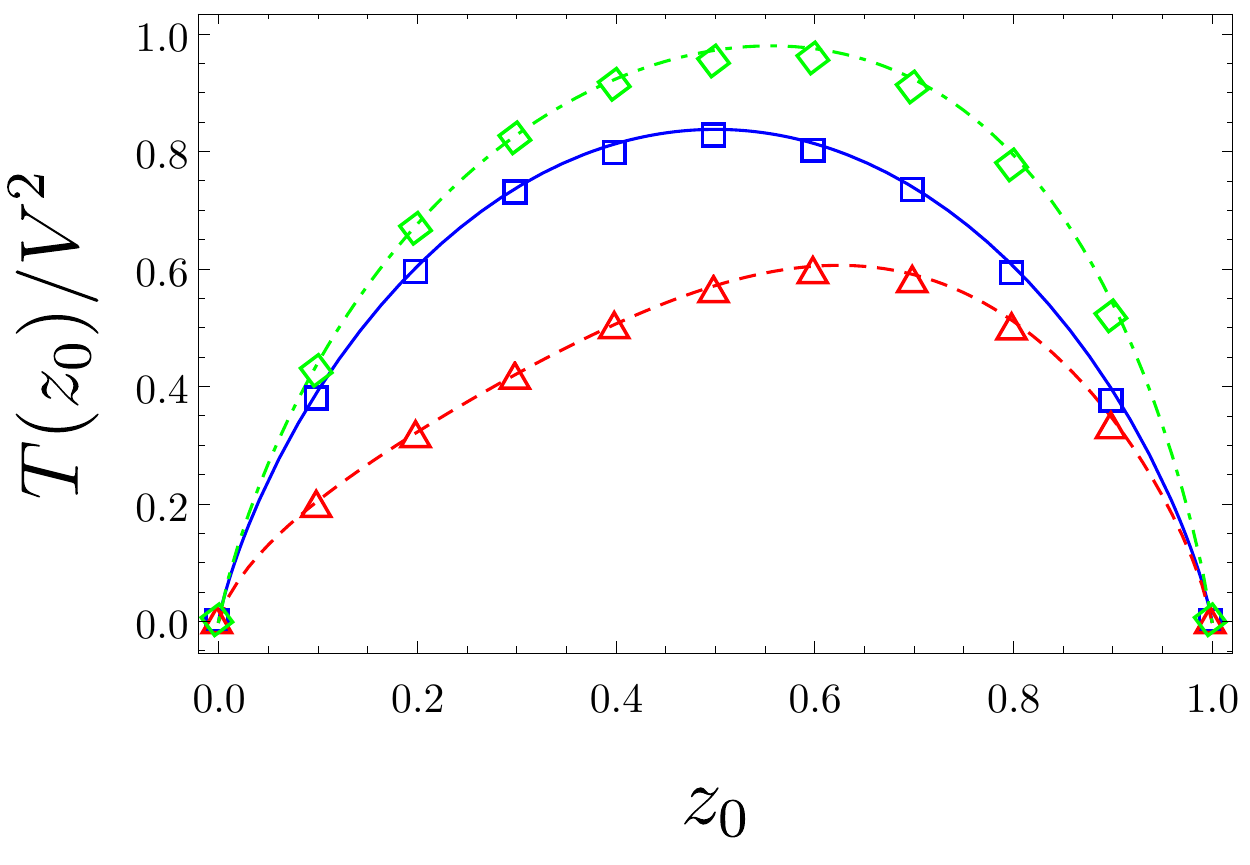}
\caption{Fixation probability of allele $1$ (left) and mean unconditional time to fixation (right) as a function of the projected initial condition $z_0$ (denoted by $z^{\rm CMIC}$ in the text) for a system with $\mathcal{D}=2$, $V=150$, and $\kappa=1.5$. Blue (squares): neutral case; red (triangles, dashed): case with selection showing an unstable internal fixed point, with $\phi_{\rm eff}^{(1)}\approx -1.33$, $\phi_{\rm eff}^{(2)}\approx -0.15$, $\Gamma_{\rm eff}\approx -1.48$, and $z^*\approx 0.9$; green (diamonds, dot-dashed): case with selection showing a stable internal fixed point, with $\phi_{\rm eff}^{(1)}\approx 0.21$, $\phi_{\rm eff}^{(2)}\approx 0.61$, $\Gamma_{\rm eff}=0.82$, and $z^*\approx 0.26$. Symbols are obtained as the mean of 20000 stochastic simulations of the microscopic system, while the lines correspond to the theoretical predictions for the fixation probability and mean time to fixation, obtained from Eqs.~(\ref{Q_0}) and (\ref{T_0}) in the neutral case, and from Eq.~(\ref{Q_s}) and the analytical solution to Eq.~(\ref{ODE_time}) in the case with selection. The value of the selection parameter is $\epsilon=0.03$.}
\label{fig:TQ_2D}
\end{figure}

%%%%%%%%%%%%%%%%%%%%%%%%%%%%%%%%%%%%%%%%%%%%%%%%%%%%%%%%%%%%%%%%%%%%%%%%%%%%%%%
%%%%%%%%%%%%%%%%%%%%%%%%%%%%%%%%%%%%%%%%%%%%%%%%%%%%%%%%%%%%%%%%%%%%%%%%%%%%%%%

To investigate this, we use the form of $\bar{A}(z)$ given by Eq.~(\ref{Abar_like_Ref_Oned}). There is the possibility of an internal fixed point at $z^* = \phi^{(1)}_{\rm eff}/\Gamma_{\rm eff}$ if $\Gamma_{\rm eff} \neq 0$, but clearly we require $0 < z^* < 1$, for this to be an internal fixed point in a biologically relevant regime. If we introduce the quantity 
\begin{equation}
\phi^{(2)}_i \equiv  \hat{c}^{(2 2)}_i - \hat{c}^{(2 1)}_i\,,
\label{phi_2_defn}
\end{equation}
in an analogous way to $\phi^{(1)}_i$, then we can easily show, as in the one-island case~\cite{constable2015a}, that if $0 < z^* < 1$, then either $\phi^{(\alpha)}_{\rm eff} > 0$ (for both $\alpha=1$ and $\alpha=2$) or $\phi^{(\alpha)}_{\rm eff} < 0$ (again for both $\alpha=1$ and $\alpha=2$). We can also investigate the stability of the internal fixed point. A simple calculation shows that the internal fixed point is stable if $\Gamma_{\rm eff} > 0$ and unstable if $\Gamma_{\rm eff} < 0$. Since $\Gamma_{\rm eff} = \phi^{(1)}_{\rm eff} + \phi^{(2)}_{\rm eff}$, an internal fixed point exists and is stable if both $\phi^{(\alpha)}_{\rm eff}$ are positive---as shown in Fig.~\ref{fig:phase_select}---and it exists and is unstable if both $\phi^{(\alpha)}_{\rm eff}$ are negative.

Two quantities which are of interest to calculate are the fixation probability of a given allele and the mean time to fixation of the system, given a set of initial allele frequencies. These are also useful to test the approximations that have been made to obtain the reduced model, since they are long-time properties in the sense that we expect fixation to occur after the system has reached the SS, and has moved along the SS to reach either $z=0$ or $z=1$.

To calculate the fixation probability and mean time to fixation, we revert to the formalism of Fokker-Planck equations. The details of the calculation are given in the SM (Sec.~\ref{sec:TQ_SM}); here we simply compare these results against simulations of the full system, shown in Fig.~\ref{fig:TQ_2D} for $\mathcal{D}=2$---and $\mathcal{D}=4$ in Fig.~\ref{fig:TQ_4D} (see SM). When there is no selection, we find that the agreement between theory and simulation is excellent. When selection is present, we also see that in spite of the relatively large values of the selection parameter explored, the calculation carried out to linear order in $\epsilon$ captures the behaviour of the full system extremely well. Furthermore, we corroborate the supposition that the existence of a stable (resp. unstable) internal fixed point of the reduced system leads to larger (resp. smaller) values of the fixation time. In Fig.~\ref{fig:TQ_2D}, we present a version of the system with $\hat{c}^{(11)}_i,\hat{c}^{(22)}_i>0$ and $\phi_i^{(1)},\phi^{(2)}_i<0$ for all $i$, so that $\phi_{\rm eff}^{(\alpha)} < 0$, $\alpha=1,2$, yielding an unstable fixed point. This is compared to a version with the signs of $\hat{c}^{(12)}_i$ and $\hat{c}^{(21)}_i$ reversed so that, all the other parameters being equal, in this case $\phi_i^{(1)},\phi^{(2)}_i>0$ for all $i$ and the fixed point is stable. The difference between both scenarios is clearly seen. A stronger effect is observed for the case with $\mathcal{D}=4$---see Fig.~\ref{fig:TQ_4D}---which shows much longer times to fixation when a stable fixed point is present.

\section{Discussion}
In this paper we have investigated a model of metapopulation genetics and shown that, despite its relative complexity, it could be reduced to an effective model with only one degree of freedom. This model is amenable to mathematical analysis.

Our starting point differed from that used by many theoretical population geneticists in so far that we did not use the Wright-Fisher or Moran model in their original microscopic form or in their mesoscopic form obtained through the diffusion limit. Although these models are widely used, they have several disadvantages. We have already mentioned the artifically fixed population size, which is required because the models do not include competition between individuals which potentially leads to a rapid increase in population size. Another example, especially relevant in this paper, is the convoluted way in which the migration process is described in the Moran model. 

In the SLVC model, individuals simply migrate at a certain rate, just as they are born, die or compete with each other at a certain rate. Therefore, in Eq.~(\ref{trans_rates}), the transition rates for migration only depend on the population density of the relevant allele on the island from which the migration takes place, $j$. As a consequence it is linear in this density, but it changes the population size on both island $j$ and on island $i$ where the migrant moves to. By contrast, in the Moran model the transition rates depend on the population density of the relevant allele on both islands. It is quadratic in the densities, although cancellations mean that eventually it turns out to be linear, but still depending on the densities of the relevant allele on both $j$ and $i$. In addition, the migration process only changes the make-up of the population on island $i$ (by perhaps displacing a resident of that island), but does not change the make-up of the population on island $j$, since all that happens here is that an offspring of an individual migrates as soon as it is born. The process then, in the SLVC model, is clearly simpler and more intuitive. A disadvantage of the SLVC model is, of course, that it doubles the number of variables, as compared to the Moran model, but it can still be reduced to an effective one-variable model, just as in the case of the Moran model~\cite{constable2014a,constable2014b}. 

The method we have discussed in this paper can be extended to SLVC models with additional features. For instance, in addition to migration, selection and genetic drift, the process of mutation could be added, as has been done for the Moran model~\cite{constable2015c}. There are however many other effects that could be included: the individuals could be assumed to be diploid, or the effect of more than one loci could be included or other types of ecological interactions could be incorporated. There would then be many types of fast modes, but as long as there was a time-scale separation between these and a few slow modes, there would be the possibility of an effective model with just a few degrees of freedom which would encapsulate the essence of the full model. In this way it may be possible to gain quantitative insights into quite complex models.

\clearpage

\def\bibsection{\section*{\refname}}
\renewcommand{\baselinestretch}{1}
\renewcommand{\thefigure}{SM\arabic{figure}}
\renewcommand{\theequation}{SM\arabic{equation}}
\renewcommand{\thesection}{\arabic{section}} 
\setcounter{section}{0}
\setcounter{equation}{0}
\setcounter{figure}{0}
\setcounter{table}{0}
\setcounter{page}{1}
\makeatletter
\renewcommand{\bibnumfmt}[1]{[#1]}
\renewcommand{\citenumfont}[1]{#1}

\centerline{\large{SUPPLEMENTARY MATERIAL}}

\vspace{0.8cm}

\centerline{\large{\bf Reduction of a metapopulation genetic model to an effective one island model}}

\vspace{0.5cm}

\centerline{C\'{e}sar Parra-Rojas$^1$ and Alan~J. McKane$^2$}

\vspace{0.3cm}

\noindent $^1$Frankfurt Institute for Advanced Studies, 60438 Frankfurt am Main, Germany\\
$^2$Theoretical Physics Division, School of Physics and Astronomy, The University of Manchester, Manchester M13 9PL, UK

\bigskip

\section{Introduction}

\medskip

The historical development of population genetics had some unusual aspects, one of which was reliance---rare in the biological sciences---on mathematical models. The ``modern synthesis''~\cite{dobzhansky1937a} started with the work of Fisher, Wright and Haldane, which was based on the analysis of simple models and played a large part in the wide acceptance of the idea of natural selection~\cite{mayr1942a}. These models were extended in subsequent years~\cite{ewens2004a}, but these developments were often divorced from advances in ecological theory~\cite{roughgarden_1979a}. Another feature was the elaboration and increasing complexity of the models: as discussed in the main text, the addition of features such as spatial structure, sexual reproduction or several gene loci, made it increasingly difficult to make analytical progress with the solution of such models. It is these two components---the detachment from ecological theory that many models of population genetics display, and the difficulty in analysing more realistic models---that underlie the objectives of this paper.

The difficulties in carrying out a mathematical analysis of models with distinct subpopulations have resulted in this area of population genetics being less well explored than many others. Very early on in the development of the subject, Wright~\cite{wright1931a} studied what is now referred to as the standard island model, although there was no actual spatial structure assumed. Much later the stepping stone model~\cite{kimuraSSMa} did contain a very simple spatial structure: a one-dimensional line of islands, with migration only allowed from an island to its nearest neighbours. A study of fixation in a model with spatial population structure by Maruyama~\cite{maruyama1970a} led to several further investigations~\cite{nagylaki1980SMa,barton1993a,whitlock2003a}; the book by Rousset~\cite{rousset2004a} gives a comprehensive review of these, and other, contributions.

The variety of models of spatial structure, the numerous approximations that were used to investigate them, and the difficulty in assessing the accuracy of the predictions, recently led us to carry out an investigation of metapopulation genetics, where the starting point was simple and clear and where the approximations were few and as generic as possible~\cite{constable2014aa,constable2014ba}. We will use a similar approach here, but using the SLVC model rather than a metapopulation version of the Moran model. The case of a single island SLVC model has been analysed previously~\cite{constable2015aa}, and the present paper can be viewed as a generalisation of this work to a model with spatial structure. A further purpose of the paper is to provide a concise review of the methodology we are using; the main text provides an overview of the method together with the key results for the specific model we investigate, while this supplementary material gives further details.

The outline of this document is as follows. In Sec.~\ref{sec:setup_SM} we set up the model in a form which is as simple as possible, if it is to capture the processes that we wish to describe. The use of the diffusion approximation allows the model to be written as a stochastic differential equation which is given in the main text. In this form the fast and slow modes of the dynamics can be identified; these are determined explicitly in Sec.~\ref{sec:slow_modes_SM}. This identification is used in Sec.~\ref{sec:reduction_SM} to derive a reduced model, which has only one degree of freedom. This is a significant simplification that allows us to calculate the probability of fixation and the mean time to fixation of the alleles. This is carried out in Sec.~\ref{sec:TQ_SM}, where the results are compared to numerical simulations of the original model.

\medskip

\section{Formulation of the model and the diffusion approximation}\label{sec:setup_SM}

\medskip

We begin the discussion of the construction and development of the model by specifying the constituents. The number of haploid individuals occupying island $i$, with $i=1,\ldots,\mathcal{D}$, which carry allele $1$ will be denoted by $n^{(1)}_i$, and the number which carry allele $2$ on the same island by $n^{(2)}_i$. They will reproduce at rates $b^{(1)}_i$ and $b^{(2)}_i$ respectively and die at rates $d^{(1)}_i$ and $d^{(2)}_i$. We will also allow for competition between individuals of type $\alpha$ and $\beta$ on island $i$, at a rate $c^{(\alpha \beta)}_i$, $\alpha,\beta=1,2$. This will tend to regulate the population size, without imposing the condition that $n^{(1)}_{i}+n^{(2)}_{i}$ is fixed on each island $i$. The processes introduced so far are local to island $i$, but we are also required to introduce migration between the islands. This is assumed to be independent of the other processes, and so we will denote by $\mu_{ij}$ the rate at which an individual from island $j$ will migrate to island $i$. This process will only be defined for $i \neq j$. Note that one could make $\mu_{ij}$ dependent on the allele type $\alpha=1,2$, however here we will assume that the migration rates for both alleles are equal. We will use the notation $\underline{n}^{(1)}=(n^{(1)}_1,\ldots, n^{(1)}_{\mathcal{D}})$, $\underline{n}^{(2)}=(n^{(2)}_1,\ldots, n^{(2)}_{\mathcal{D}})$, and $\bm{n}=(\underline{n}^{(1)},\underline{n}^{(2)})$ to describe the occupation numbers of the system concisely.

The state of the system, $\bm{n}$, will change according to whether individuals of type $1$ or type $2$ on the various islands change due to one or more of the above processes. To define the dynamics of the system, we need to give the rate of transition from the current state, $\bm{n}$, to a new state $\bm{n}'$. These are taken to be
\begin{eqnarray}
\label{trans_rates}
& & T_{1,i}(n^{(1)}_i + 1,n^{(2)}_i|n^{(1)}_i,n^{(2)}_i) =
b^{(1)}_i\frac{n^{(1)}_i}{V_i}, \nonumber \\
& & T_{2,i}(n^{(1)}_i,n^{(2)}_i + 1|n^{(1)}_i,n^{(2)}_i) =
b^{(2)}_i\frac{n^{(2)}_i}{V_i}, \nonumber \\
& & T_{3,i}(n^{(1)}_i - 1,n^{(2)}_i|n^{(1)}_i,n^{(2)}_i) =
d^{(1)}_i\frac{n^{(1)}_i}{V_i} 
+ c^{(11)}_{i}\frac{n^{(1)}_i}{V_i}\frac{n^{(1)}_i}{V_i} +  
c^{(12)}_{i}\frac{n^{(2)}_i}{V_i}\frac{n^{(1)}_i}{V_i}, \\
& & T_{4,i}(n^{(1)}_i,n^{(2)}_i - 1|n^{(1)}_i,n^{(2)}_i) =
d^{(2)}_i\frac{n^{(2)}_i}{V_i} 
+ c^{(22)}_{i}\frac{n^{(2)}_i}{V_i}\frac{n^{(2)}_i}{V_i} 
+ c^{(21)}_{i}\frac{n^{(1)}_i}{V_i}\frac{n^{(2)}_i}{V_i}, \nonumber \\ 
& & T_{5, i j}(n^{(1)}_i + 1,n^{(2)}_i,n^{(1)}_j - 1,n^{(2)}_j|\bm{n}) =
\mu_{ij} \frac{n^{(1)}_j}{V_j}, \ (i \neq j), \nonumber \\
& & T_{6, i j}(n^{(1)}_i,n^{(2)}_i + 1,n^{(1)}_j,n^{(2)}_j - 1|\bm{n}) =
\mu_{ij} \frac{n^{(2)}_j}{V_j}, \ (i \neq j), \nonumber
\end{eqnarray}
where in the arguments of the rates we only list those variables that are involved in the reaction and where the initial state is given on the right and the final state on the left. Here, $T_{1,i}$ (resp. $T_{2,i}$) corresponds to the birth of an individual of type $1$ (resp. $2$) on island $i$; $T_{3,i}$ (resp. $T_{4,i}$) corresponds to the death, either natural or due to competition, of an individual of type $1$ (resp. $2$) on island $i$; and $T_{5,ij}$ (resp. $T_{6,ij}$) corresponds to the migration of an individual of type $1$ (resp. $2$) from island $j$ to island $i$.

The transition rates given by Eq.~(\ref{trans_rates}) are those which give Lotka-Volterra competition equations in the deterministic limit and we therefore describe them as defining the SLVC metapopulation model. The migration process is the simplest possible, and therefore taken together these are arguably the simplest stochastic dynamics which encodes the processes that we wish to include in the model. They are also a generalisation of the SLVC model on one island, which was studied previously~\cite{constable2015aa}. The factors $V_i$ denote the potential capacity of island $i$, both in terms of environmental factors required to sustain a population and the size of the island. As such, they are the carrying capacity of each island, but without the sense of a sharp cut-off, but rather give a soft cut-off. We will assume that the carrying capacities of the islands vary among them---some can be more fertile or larger than others---but not by orders of magnitude. Therefore we will set $V_i = \beta_i V$, where $\beta_i$ is a number of $\mathcal{O}(1)$ that characterises the capacity of each island compared to the others, and where $V$ is the typical carrying capacity of an island, which will be used in the application of the diffusion approximation.

The transition rates describe how the system changes in an infinitesimal time step during which one particular process occurs. To describe the stochastic dynamics over a finite time-interval we need to introduce a differential equation that describes how the probability distribution function of the system in state $\bm{n}$, $P(\bm{n},t)$, changes in time due to these transitions. This is the master equation, which takes the generic form~\cite{vanKampen2007a}
\begin{equation}
\frac{\mathrm{d}P(\bm{n},t)}{\mathrm{d}t} = \sum_{\bm{n}' \neq \bm{n}}\left[ T(\bm{n}|\bm{n}')P(\bm{n}',t) - T(\bm{n}'|\bm{n})P(\bm{n},t)\right],
\label{master_generic}
\end{equation}
where the transition rate $T(\bm{n}'|\bm{n})$ represents all the transitions rates given in Eq.~(\ref{trans_rates}). 

The master equation (\ref{master_generic}) can be expressed more fully by writing the right-hand side of Eq.~(\ref{master_generic}) as 
\begin{equation}
\sum^6_{\mu=1} \left\{ \sum_{\bm{n}' \neq \bm{n}}\left[ T_\mu(\bm{n}|\bm{n}')P(\bm{n}',t) - T_\mu(\bm{n}'|\bm{n})P(\bm{n},t)\right] \right\},
\label{master_basic}
\end{equation}
where the sum on $\mu$ is a sum over the six distinct types of transitions rates listed in Eq.~(\ref{trans_rates}). We can go further, and specify the transition rates as they are given in Eq.~(\ref{trans_rates}) by writing out the master equation in terms of what are in effect stoichiometric coefficients, which tell us how many individuals are transformed to other forms or to other islands by the ``reactions'' $\mu=1,\ldots,6$. In the notation introduced above for the master equation, $\bm{n}'=\bm{n}-\bm{\nu}$, where we will write $\bm{\nu}_\mu$ for the stoichiometric vector corresponding to reaction $\mu$. Specifically the master equation now takes the form
\begin{eqnarray}
& & \frac{\mathrm{d}P(\bm{n},t)}{\mathrm{d}t} = \sum^{4}_{\mu=1} 
\sum^{\mathcal{D}}_{i=1} \left[ T_{\mu,i}(\bm{n}|\bm{n}-\bm{\nu}_{\mu,i})P(\bm{n}-\bm{\nu}_{\mu,i},t) - T_{\mu,i}(\bm{n}+\bm{\nu}_{\mu,i}|\bm{n})P(\bm{n},t) \right] 
\nonumber \\
& & + \sum^{6}_{\mu=5} \sum^{\mathcal{D}}_{i=1} \sum^{\mathcal{D}}_{j \neq i} 
\left[ T_{\mu,i j}(\bm{n}|\bm{n}-\bm{\nu}_{\mu, i j})
P(\bm{n}-\bm{\nu}_{\mu, i j},t) - T_{\mu,i j}(\bm{n}+\bm{\nu}_{\mu, i j}|\bm{n})P(\bm{n},t) \right],
\label{master_alt}
\end{eqnarray}
where $\bm{\nu}_{\mu,i}$ describes how many individuals on island $i$ are transformed during the reactions $\mu=1,\ldots,4$ and $\bm{\nu}_{\mu,i j}$ describes how many individuals on islands $i$ and $j$ are transformed during the reactions $\mu=5,6$. The specific forms of the $\bm{\nu}_{\mu, i}$ and $\bm{\nu}_{\mu, i j}$ are:
\begin{eqnarray}
\bm{\nu}_{1, i} &=& (0,\ldots,1,0,\ldots,0) \ \ (\textrm{non-zero\ entry\ at\ }i), \nonumber \\
\bm{\nu}_{2, i} &=& (0,\ldots,1,0,\ldots,0) \ \ (\textrm{non-zero\ entry\ at\ }\mathcal{D}+i), \nonumber \\
\bm{\nu}_{3, i} &=& (0,\ldots,-1,0,\ldots,0) \ \ (\textrm{non-zero\ entry\ at\ }i), \nonumber \\
\bm{\nu}_{4, i} &=& (0,\ldots,-1,0,\ldots,0) \ \ (\textrm{non-zero\ entry\ at\ }\mathcal{D}+i), \nonumber \\
\bm{\nu}_{5, i j} &=& (0,\ldots,1,0,\ldots,-1,\ldots,0), \nonumber \\
\bm{\nu}_{6, i j} &=& (0,\ldots,1,0,\ldots,-1,\ldots,0),
\label{stoichiometric}
\end{eqnarray}
where in the last two cases the entry $1$ ($-1$) is at position $i$ ($j$) for $\bm{\nu}_{5, i j}$ and at position $\mathcal{D}+i$ ($\mathcal{D}+j$) for $\bm{\nu}_{6, i j}$, where $i \neq j$.

The master equation (\ref{master_alt}), together with the transition rates in Eq.~(\ref{trans_rates}) and an initial condition for $P(\bm{n},t)$, gives a complete description of the stochastic dynamics of the system. It is this basic form that is used in numerical simulations later in the paper. 

While the form of the master equation (\ref{master_alt}) appears to be far more complicated than the master equation (\ref{master_basic}), it has the great advantage that the first approximation that is used to simplify this rather complicated dynamics---as described in the main text---can be applied in an almost algorithmic fashion. This is the diffusion approximation, where it is assumed that the $V_i$ are sufficiently large so that $x^{(\alpha)}_i \equiv n^{(\alpha)}_i/V_i$, $i=1,\ldots,\mathcal{D}$, $\alpha=1,2$, are approximately continuous. This is a large-$V$ approximation~\cite{crow2009a}, and so another aspect of the approximation is to expand the master equation as a power series in $V^{-1}$ to obtain the Fokker-Planck equation~\cite{gardiner2009a,risken1989a}. Before giving this equation, however, we describe some notation to make it look a little simpler: we introduce an index $I$ that runs from $1$ to $2\mathcal{D}$, so that $I=i$ if the allele labelled is $1$ and if the island being considered is $i$, and $I=\mathcal{D}+i$ if the allele labelled is $2$ and if the island being considered is $i$. Then the Fokker-Planck equation takes the form
\begin{equation}
\frac{\partial P(\bm{x},t)}{\partial t} =
- \frac{1}{V}\,\sum_{I=1}^{2\mathcal{D}} \frac{\partial }{\partial x_I} 
\left[ A_{I}(\bm{x}) P(\bm{x},t) \right] + 
\frac{1}{2V^2} \sum_{I,J=1}^{2\mathcal{D}} \frac{\partial^2 }{\partial x_I \partial x_J} \left[ B_{I J}(\bm{x}) P(\bm{x},t) \right], 
\label{FPE_full}
\end{equation}
where we have neglected terms of order $V^{-3}$ and higher, and where $\bm{x} = (\underbar{x}_1,\underbar{x}_2) = (x^{(1)}_1,x^{(2)}_1,\ldots,$ $x^{(\mathcal{D})}_1,x^{(1)}_2,x^{(2)}_2,\ldots,x^{(\mathcal{D})}_2)$.

Equation (\ref{FPE_full}) is simply a generic Fokker-Planck equation; we require to derive the specific forms for the functions $A_I(\bm{x})$ and $B_{IJ}(\bm{x})$ for the model under consideration. In Ref.~\cite{mckane2014a} it is shown that performing the diffusion approximation, that is going over to the continuous variables $\bm{x}$, and expanding the master equation in powers of $V^{-1}$, gives the Fokker-Planck equation with the functions $A_I(\bm{x})$ and $B_{IJ}(\bm{x})$ given as explicit sums over the reactions $\mu$ with stoichiometric coefficients $\bm{\nu}_\mu$. In this way Eqs.~\eqref{full_A}--\eqref{rest_of_B} of the main text can be obtained directly from Eqs.~(\ref{trans_rates})--(\ref{stoichiometric}). The functions $A_I(\bm{x})$ and $B_{I J}(\bm{x})$ specify the model and are derived from, and are in effect the continuous versions of, the transition rates given in Eq.~(\ref{trans_rates}). 

As we discuss below, $A^{(\alpha)}_i(\bm{x})$ is the only function that appears in the deterministic description. It consists of the familiar Lotka-Volterra local terms involving birth, death and competition of the $\alpha$ allele on island $i$, together with the migration of this allele between island $i$ and the other islands, as described by the term $\mathcal{M}^{(\alpha,-)}_i$. The $B_{IJ}(\bm{x})$ only appear in the stochastic dynamics. As mentioned in the main text, the content of the Fokker-Planck equation can be written in a more intuitive way, in the form of the equivalent It\={o} stochastic differential equation. The Fokker-Planck equation (\ref{FPE_full}) or alternatively Eqs.~\eqref{SDE_full} and~\eqref{correlator} together give the mesoscopic description of the system. The familiar, deterministic, Lotka-Volterra equations (together with migration) form the macroscopic description, and can be found by taking the $V \to \infty$ limit of Eq.~\eqref{SDE_full}.

The chief virtue of the diffusion approximation is to move away from discrete variables to continuous ones, which are easier to analyse. However, as is typically the case when spatial structure is introduced, even the continuous form of the model is not easy to study, here exemplified by the complicated nature of the $A_I(\bm{x})$ and $B_{I J}(\bm{x})$. We therefore now move on to discuss a second approximation, which will have the effect of reducing the model to a one-dimensional effective theory, which can nevertheless make accurate predictions about the original form of the model.

\medskip

\section{Model reduction I.~Identification of the slow and fast modes}\label{sec:slow_modes_SM}

\medskip

In this section we will give further details of the identification of the slow and fast modes of the original mesoscopic model. We begin with the model with the selection parameter, $\epsilon$, set equal to zero. The preliminary analysis is given in the main text, where it is shown that a CM exists which is given by $y^{(\alpha)}_i = y^{(\alpha)}$ for all $i$ and $\alpha=1,2$, with $y^{(1)} + y^{(2)} = 1$ (see Eq.~\eqref{CM_y}), where the $y^{(\alpha)}_i$ are scaled versions of the original parameters $x^{(\alpha)}_i$. We chose the CM to be parameterised by $y^{(1)}$ which we denote by $z$, the only variable of the reduced system.

Further insight can be gained by calculating the Jacobian on the CM. To do this, we first write the deterministic equation for $y^{(\alpha)}_i$, analogous to Eq.~(\ref{neutral_deterministic}) for $x^{(\alpha)}_i$. This is given by
\begin{eqnarray}
\frac{\mathrm{d}y^{(\alpha)}_i}{\mathrm{d}\tau} &=& \frac{\kappa c^{(0)}_i y^{(\alpha)}_i}{\beta_i} \left\{  1 - \left[ y^{(1)}_i + y^{(2)}_i \right] \right\} + \sum^{\mathcal{D}}_{j=1} H_{ij} y^{(\alpha)}_j, \ \ \ i=1,\ldots,\mathcal{D}, \ \ \alpha=1,2,
\label{y_neutral_deterministic}
\end{eqnarray}
where
\begin{equation}
H_{ij} = \frac{\mu_{ij}}{\beta_i}, \ \ \ \mathrm{if}\ i \neq j,\ \ \ \ 
H_{ii} = - \sum^{\mathcal{D}}_{j \neq i} \frac{\mu_{ij}}{\beta_i}.
\label{H_defn}
\end{equation}
Differentiating the right-hand side of Eq.~(\ref{y_neutral_deterministic}) by $y^{(\beta)}_k$ and setting $y^{(1)}_k = z$ and $y^{(2)}_k = 1 - z$, one obtains the Jacobian
\begin{equation}
J = \left( \begin{array}{cc} 
\mathcal{J} z + H & \ \ \ \mathcal{J} z \\ \\
\mathcal{J}  (1-z) & \ \ \ \mathcal{J} (1-z) + H
\end{array}
\right)\,,
\label{Jacobian}
\end{equation}
where $\mathcal{J}$ is a $\mathcal{D}$-dimensional diagonal matrix with entries given by $\mathcal{J}_{ij} = - (c^{(0)}_i \kappa/\beta_i) \delta_{ij}$. 

%%%%%%%%%%%%%%%%%%%%%%%%%%%%%%%%%%%%%%%%%%%%%%%%%%%%%%%%%%%%%%%%%%%%%%%%%%%
%%%%%%%%%%%%%%%%%%%%%%%%%%%%%%%%%%%%%%%%%%%%%%%%%%%%%%%%%%%%%%%%%%%%%%%%%%%

\begin{figure}
\centering
%\includegraphics[width=0.8\columnwidth]{fig2b.pdf}\\
%\ \\
\includegraphics[width=0.48\columnwidth]{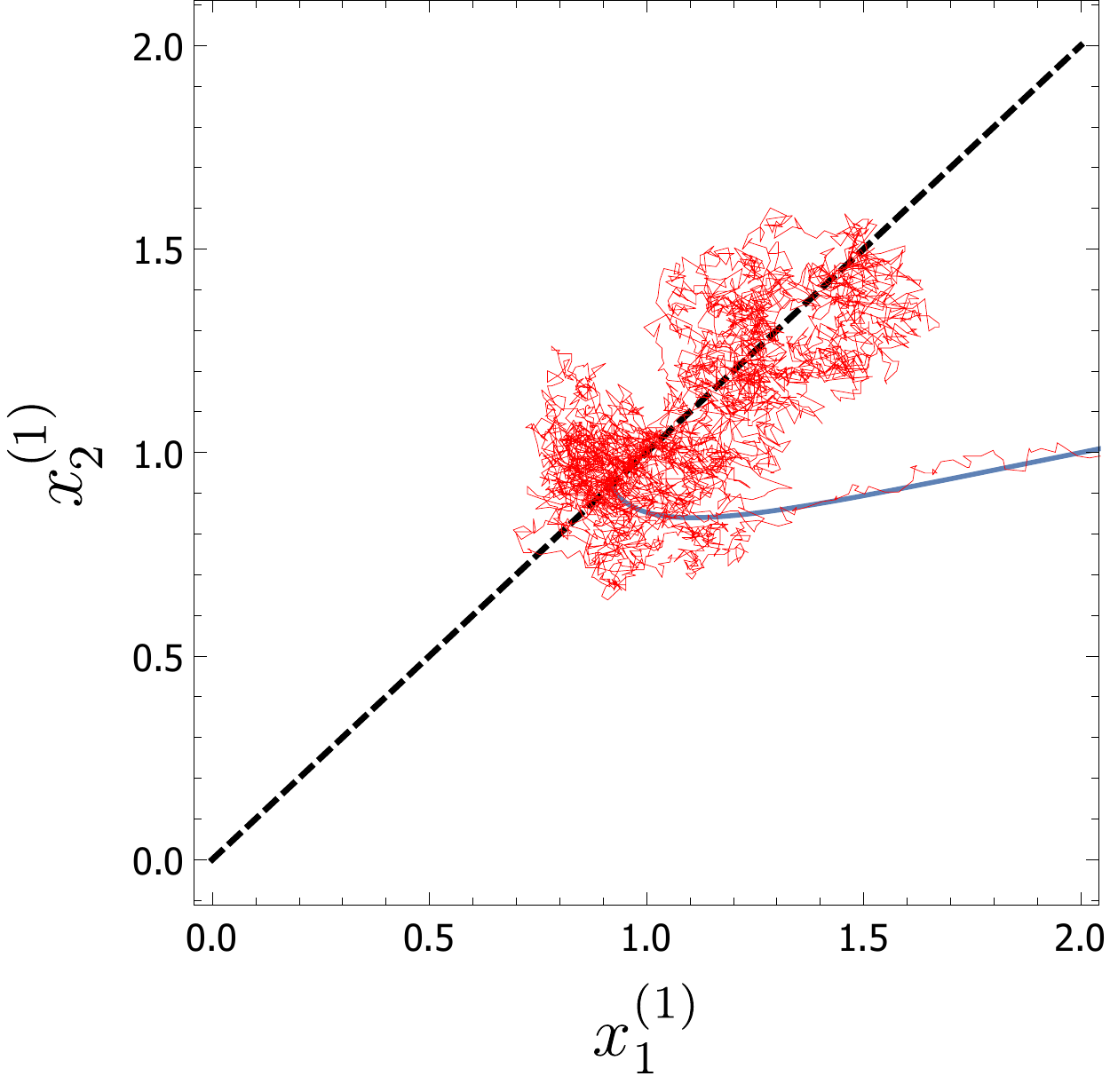}
\caption{A neutral system with two islands. Phase diagram for individuals of type $1$ on islands $1$ and $2$. Blue line: deterministic trajectory; red line: one stochastic trajectory; black, dashed line: neutral solution, $x_1^{(\alpha)}=x_2^{(\alpha)}$, $\alpha=1,2$. Parameters: $V=300$, $\kappa =1.5$.}
\label{fig:phase_neutral_SM}
\end{figure}

%%%%%%%%%%%%%%%%%%%%%%%%%%%%%%%%%%%%%%%%%%%%%%%%%%%%%%%%%%%%%%%%%%%%%%%%%%%
%%%%%%%%%%%%%%%%%%%%%%%%%%%%%%%%%%%%%%%%%%%%%%%%%%%%%%%%%%%%%%%%%%%%%%%%%%%

We will now give details of the nature of the eigenvalues, and the structure of the eigenvectors, of the Jacobian, $J$, defined by Eq.~(\ref{Jacobian}).

We begin the analysis by recalling the form of the eigenvectors in the one island case~\cite{constable2015aa}:
\begin{eqnarray}\label{eigenvectors_one_island}
\bm{u}^{\{ {\rm one}\} } &=& \left( \begin{array}{c} 1 - z \\ - z \end{array} \right), \ \ \bm{v}^{\{ {\rm one}\} } = \left( \begin{array}{c} 1 \\ - 1 \end{array} \right); \nonumber \\
\bm{u}^{\{ {\rm two}\} } &=& \left( \begin{array}{c} 1 \\ 1 \end{array} \right), \ \ 
\bm{v}^{\{ {\rm two}\} } = \left( \begin{array}{c} z \\ 1 - z \end{array} \right).
\end{eqnarray}  
Motivated by these we will now show that the eigenvectors of the Jacobian (\ref{Jacobian}) fall into the two classes
\begin{eqnarray}
\label{eigenvectors_general_form}
\left( \begin{array}{c} (1 - z)\underline{\alpha}_{\rm L} \\ - z \underline{\alpha}_{\rm L} \end{array} \right), \left( \begin{array}{c} \underline{\alpha}_{\rm R} \\ - \underline{\alpha}_{\rm R} \end{array} \right);
\ \ \left( \begin{array}{c} \underline{\beta}_{\rm L} \\ \underline{\beta}_{\rm L} \end{array} \right), \left( \begin{array}{c} z \underline{\beta}_{\rm R} \\ (1 - z)\underline{\beta}_{\rm R} \end{array} \right).
\end{eqnarray} 
The proof is very simple, and just consists of applying the Jacobian matrix to the eigenvectors in Eq.~(\ref{eigenvectors_general_form}). One finds that they are indeed eigenvectors, as long as the $\underline{\alpha}s$ and $\underline{\beta}s$ obey the equations
\begin{eqnarray}
& & \underline{\alpha}_{\rm L}H = \lambda\underline{\alpha}_{\rm L}, \ \ 
H\underline{\alpha}_{\rm R} = \lambda\underline{\alpha}_{\rm R}; \ \ 
\underline{\beta}_{\rm L} \left( H + \mathcal{J} \right) = 
\lambda\underline{\beta}_{\rm L}, \nonumber \\
& & \left( H + \mathcal{J} \right) \underline{\beta}_{\rm R} = \lambda\underline{\beta}_{\rm R},
\label{eigen_conditions}
\end{eqnarray}
where $\lambda$ is a constant. That is, $\underline{\alpha}_{\rm L}$ and $\underline{\alpha}_{\rm R}$ are left- and right-eigenvectors of $H$ respectively, and $\underline{\beta}_{\rm L}$ and $\underline{\beta}_{\rm R}$ are left- and right-eigenvectors of $H + \mathcal{J}$ respectively. Since these are $2\mathcal{D}$ eigenvectors, which are assumed independent, we have reduced finding the eigenvalues and eigenvectors of $J$ to finding the eigenvalues and eigenvectors of (i) $H$, and (ii) $H + \mathcal{J}$.

Let us denote the eigenvectors as follows:
\begin{eqnarray}
\bm{U}^{\{ I\} } &=& \left( \begin{array}{c} (1 - z)\underline{\alpha}_{\rm L} \\ - z \underline{\alpha}_{\rm L} \end{array} \right); \ 
\bm{V}^{\{ I\} } = \left( \begin{array}{c} \underline{\alpha}_{\rm R} \\ - \underline{\alpha}_{\rm R} \end{array} \right), \ I \leq \mathcal{D}\,,
\nonumber \\
\bm{U}^{\{ I\} } &=& \left( \begin{array}{c} \underline{\beta}_{\rm L} \\ \underline{\beta}_{\rm L} \end{array} \right), \ 
\bm{V}^{\{ I\} } = \left( \begin{array}{c} z \underline{\beta}_{\rm R} \\ (1 - z)\underline{\beta}_{\rm R} \end{array} \right), \ I \geq \mathcal{D}\,.
\label{eigenvectors}
\end{eqnarray}
The orthonormality properties of the eigenvectors (\ref{eigenvectors}) follow from those for the $\underline{\alpha}s$ and $\underline{\beta}s$, since
\begin{eqnarray}
\bm{U}^{\{ I\} {\rm T}}\cdot\bm{V}^{\{ J\} } &=& \underline{\alpha}_{\rm L}^{\rm T}\cdot\underline{\alpha}_{\rm R}, \ \mathrm{if\ } I \leq \mathcal{D}; \ J \leq \mathcal{D}\,, \nonumber \\
\bm{U}^{\{ I\} {\rm T}}\cdot\bm{V}^{\{ J\} } &=& 0, \ \mathrm{if\ } I \leq \mathcal{D}; J \geq \mathcal{D}\,, \nonumber \\
\bm{U}^{\{ I\} {\rm T}}\cdot\bm{V}^{\{ J\} } &=& 0, \ \mathrm{if\ } J \leq \mathcal{D}; I \geq \mathcal{D}\,, \nonumber \\
\bm{U}^{\{ I\} {\rm T}}\cdot\bm{V}^{\{ J\} } &=& \underline{\beta}_{\rm L}^{\rm T}\cdot\underline{\beta}_{\rm R}, \ \mathrm{if\ } I \geq \mathcal{D}; J \geq \mathcal{D}\,.
\label{orthonormality}
\end{eqnarray}
So if the $\underline{\alpha}s$ and $\underline{\beta}s$ are orthonormal, then $\sum^{2\mathcal{D}}_{K=1}\,U^{\{ I\} }_K V^{\{ J\} }_K = \delta_{I J}$.

We will occasionally denote $\underline{\alpha}_{\rm L}$ and $\underline{\alpha}_{\rm R}$ as $\underline{u}$ and $\underline{v}$ respectively, since they are the left- and right-eigenvectors of $H$. That is,
\begin{equation}
\bm{U}^{\{ i\} } = \left( \begin{array}{c} (1-z)\underline{u}^{\{ i\} } \\  - z\underline{u}^{\{ i\} } \end{array} \right), \ \ \bm{V}^{\{ i\} } = \left( \begin{array}{c} \underline{v}^{\{ i\} } \\ - \underline{v}^{\{ i\} } \end{array} \right)\,,
\label{first_D_eigenvectors}
\end{equation}
where $i=1,\ldots,\mathcal{D}$. 

From Eq.~(\ref{H_defn}) we observe that $\sum_{j=1}^{\mathcal{D}}\,H_{ij} = 0$, for all $i$. We may write this condition as the eigenvalue equation $\sum_{j=1}^{\mathcal{D}}\,H_{ij}\,v^{\{ 1\} }_j = 0$, which implies that $v^{\{ 1\} }_j =1\,\forall j$ is a right-eigenvector of $H$ with eigenvalue zero. The other eigenvalues do not have a simple form, and will be complex in general, since $\mu_{ij}$ will typically not be symmetric. However we can show that their real parts will always be negative. The proof of this statement is essentially a generalisation of that given in Sec.~III of Ref.~\cite{constable2014aa}, which we begin by recapping for convenience. 

The proof consists of introducing a matrix $\mathcal{R}$ with elements given by $\mathcal{R}_{ij}= \beta_{\rm min}H_{ij}/(\mathcal{D}-1)\mu_{\rm max}$, where $\beta_{\rm min}$ is the smallest element of the set $\{ \beta_1,\ldots,\beta_{\mathcal{D}}\}$ and $\mu_{\rm max}$ is the largest migration rate. Then, by construction, every off-diagonal element of $\mathcal{R}$ lies in the interval $(0,1]$ and every diagonal element lies in the interval $[-1,0)$. Therefore the quantities $S_{ij} \equiv \mathcal{R}_{ij} + \delta_{ij}$ are all non-negative and moreover $\sum^{\mathcal{D}}_{j=1} S_{ij} = 1$. This implies that the matrix $S$, with entries $S_{ij}$, is a stochastic matrix~\cite{gantmacher1959a,cox1965a}. Such matrices have a single largest eigenvalue equal to $1$ (if, as we have assumed, no islands are completely isolated) with all the others having a magnitude less than $1$~\cite{gantmacher1959a,cox1965a}, which implies that they have real parts which are less than $1$. Since $S$ and $\mathcal{R}$ share the same eigenvectors, with the eigenvalues of $\mathcal{R}$ being those of $S$ minus $1$, the real part of the eigenvalues of $\mathcal{R}$ are negative, apart from the largest, which is zero.

A similar argument can be made for the matrix $H + \mathcal{J}$. Here we form
\begin{equation}
\mathcal{P}_{ij} = \left\{ \left[ \frac{(\mathcal{D}-1)}{\beta_{\rm min}} \right] \left( \mu_{\rm max} + c^{(0)}_{\rm max}\kappa \right) \right\}^{-1}\left( H_{ij} + \mathcal{J}_{ij} \right),
\label{curly_P} 
\end{equation}
where $c^{(0)}_{\rm max}$ is the largest member of the set $\{ c^{(0)}_i\ : \, i=1,\ldots,\mathcal{D}\}$. Then again, by construction, every off-diagonal element of $\mathcal{P}$ lies in the interval $(0,1]$ and every diagonal element lies in the interval $[-1,0)$. We can again define $S_{ij} = \mathcal{P}_{ij} + \delta_{ij}$, and so obtain a non-negative matrix, all of whose entries are less than or equal to $1$. The difference now is that the sum of the entries of the columns of the matrix will not in general equal $1$. In fact, 
\begin{equation}
\sum^{\mathcal{D}}_{j=1} \mathcal{P}_{ij} = - \left\{ \left[ \frac{(\mathcal{D}-1)}{\beta_{\rm min}} \right] \left( \mu_{\rm max} + c^{(0)}_{\rm max}\kappa \right) \right\}^{-1} \frac{c^{(0)}_i\kappa}{\beta_i},
\label{column_sum}
\end{equation}
since $\sum_j H_{ij} = 0$ and $\mathcal{J}$ is diagonal. From Eq.~(\ref{column_sum}), $\sum^{\mathcal{D}}_{j=1} \mathcal{P}_{ij} < 0$, which implies that $\sum^{\mathcal{D}}_{j=1} \mathcal{S}_{ij} < 1$. for all $i$. From the Perron-Frobenius theorem, the largest eigenvalue of $S$ is real, positive, and is less than the maximum value of $\sum^{\mathcal{D}}_{j=1} \mathcal{S}_{ij}$ taken over all $i$~\cite{gantmacher1959a}. If we choose this eigenvalue to be $\lambda^{\{ \mathcal{D}+1\}}$, then we have that $\lambda^{\{ \mathcal{D}+1\} } < 1$. The Perron-Frobenius theorem also states that all the other (generally complex) eigenvalues of $S$ will have a magnitude less than $\lambda^{\{ \mathcal{D}+1\}}$, i.e.,~less than $1$. Therefore by the same argument as used for $H$, the real parts of the eigenvalues of $H + \mathcal{J}$ are negative.

In fact, the inequality used on Eq.~(\ref{column_sum}) can be slightly strengthened:
\begin{equation}
\sum^{\mathcal{D}}_{j=1} \mathcal{P}_{ij} \leq  - \left\{ \left[ \frac{(\mathcal{D}-1)}{\beta_{\rm min}} \right] \left( \mu_{\rm max} + c^{(0)}_{\rm max}\kappa \right) \right\}^{-1} \frac{c^{(0)}_{\rm min}\kappa}{\beta_{\rm max}},
\label{column_sum_inequality}
\end{equation}
where $c^{(0)}_{\rm min}$ is the smallest member of the set $\{ c^{(0)}_i\ : \, i=1,\ldots,\mathcal{D}\}$ and $\beta_{\rm max}$ is the largest element of the set $\{ \beta_1,\ldots,\beta_{\mathcal{D}}\}$. This implies that the real part of all the eigenvalues of $S$ are less than 
\[
1 - \left\{ \left[ \frac{(\mathcal{D}-1)}{\beta_{\rm min}} \right] \left( \mu_{\rm max} + c^{(0)}_{\rm max}\kappa \right) \right\}^{-1} \frac{c^{(0)}_{\rm min}\kappa}{\beta_{\rm max}},
\]
and so 
\begin{equation}
\Re\left[\lambda^{\{ I \} }\right]  <  - \frac{c^{(0)}_{\rm min}\kappa}{\beta_{\rm max}}, \ \ \ \ 
I=\mathcal{D}+1,\ldots,2\mathcal{D}.
\label{eigenvalue_bound}
\end{equation}

The above analysis of the eigenvalues and eigenvectors of $J$ shows that, as expected, there is a single eigenvalue equal to zero, reflecting the existence of the one-dimensional CM. The right-eigenvector corresponding to this eigenvalue points along the CM. If we assume that none of the islands are isolated, that is, there is always a sequence of non-zero migration rates connecting one island to any of the others, then we can show that the real part of all the other eigenvalues is negative. These are the $2\mathcal{D}-1$ fast modes that collapse relatively quickly, taking the system to the CM. 

To make this more concrete, we denote the right (left) eigenvectors of $J$ by $\bm{V}^{\{ I\} }$ ($\bm{U}^{\{ I\} }$) and the corresponding eigenvalues by $\lambda^{\{ I\} }$, where $I=1,\ldots,2\mathcal{D}$ (as above). We will choose the zero eigenvalue and the associated eigenvectors to be those labelled by $I=1$. In the deterministic limit of the neutral model, discussed above, the system collapses onto the CM, at which point it ceases to change, since the CM lies along the vector $\bm{V}^{\{ 1\} }$ which has eigenvalue zero. To find the position on the CM to which the system collapses we introduce the projection operator
\begin{equation}
P_{IJ} = \frac{V^{\{ 1\} }_I U^{\{ 1\} }_J}{\sum^{2\mathcal{D}}_{K=1}V^{\{ 1\} }_K  U^{\{ 1\} }_K},
\label{proj_op}
\end{equation}
which is simply equal to $V^{\{ 1\} }_I U^{\{ 1\} }_J$, using the orthonormality conditions discussed above (Eq.~\eqref{orthonormality}). Application of $P_{IJ}$ to a function containing the vector $V^{\{ I\} }_J$ will wipe out all contributions with $I \neq 1$, and leave contributions with $I=1$ unchanged. Applying it to the initial value of $\bm{y}$ set at $t=0$, which we will denote by $\bm{y}^{\rm IC}$, gives the point on the CM, discussed above, to which the system deterministically collapses to:
\begin{equation}
y^{\rm CMIC}_I = \sum^{2\mathcal{D}}_{J=1} P_{IJ} y^{\rm IC}_J = V^{\{ 1\} }_I
\sum^{2\mathcal{D}}_{J=1} U^{\{ 1\} }_J  y^{\rm IC}_J,
\label{initial_condition}
\end{equation}
where the superscript CMIC denotes `CM initial condition'. In terms of the $z$ coordinate on the CM, $z=y^{(1)}_i$, introduced earlier, this reads
\begin{equation}
z^{\rm CMIC} = \sum^{2\mathcal{D}}_{J=1} U^{\{ 1\} }_J  y^{\rm IC}_J,
\label{z_initial_condition}
\end{equation}
since $V^{\{ 1\} }_I = 1$ for $I \leq \mathcal{D}$.

Finally, we can add selection, with the equation of the SS now assumed to have the form given by Eq.~(\ref{SS_coords}) of the main text, where $Y^{(1)}_i$ and $Y^{(2)}_i$ are to be determined. Substituting these coordinates into the expressions for $A^{(1)}_{i}(\bm{y})$ and $A^{(2)}_{i}(\bm{y})$ (see Eq.~(\ref{full_A})), but restricted to the SS, gives
\begin{eqnarray}
\left. A^{(1)}_i(\bm{y})\right|_{\rm SS} &=& - \epsilon \frac{c^{(0)}_i}{\beta_i}\,\kappa z \left[ Y^{(1)}_i + Y^{(2)}_i \right]  + \epsilon \sum^{\mathcal{D}}_{j=1} H_{i j} Y^{(1)}_j \nonumber \\
&+&  \frac{\epsilon}{\beta_i} z \left\{ \left( b^{(0)}_i \hat{b}^{(1)}_i - d^{(0)}_i \hat{d}^{(1)}_i \right) - c^{(0)}_{i} \hat{c}^{(1 1)}_i \kappa z - c^{(0)}_{i} \hat{c}^{(1 2)}_i \kappa \left( 1 - z \right)\right\} + \mathcal{O}\left( \epsilon^2 \right), \nonumber \\
\left. A^{(2)}_i(\bm{y})\right|_{\rm SS} &=& - \epsilon \frac{c^{(0)}_i}{\beta_i}\,\kappa \left( 1 - z \right)\left[ Y^{(1)}_i + Y^{(2)}_2 \right] + \epsilon \sum^{\mathcal{D}}_{j=1} H_{i j} Y^{(2)}_j \nonumber \\
&+& \frac{\epsilon}{\beta_i} \left( 1 - z \right)\,\left\{ \left( b^{(0)}_i \hat{b}^{(2)}_i - d^{(0)}_i \hat{d}^{(2)}_i \right) - c^{(0)}_{i} \hat{c}^{(2 2)}_i \kappa \left( 1 - z \right) - c^{(0)}_i \hat{c}^{(2 1)}_i \kappa z \right\} + \mathcal{O}\left( \epsilon^2 \right). \nonumber \\
\label{A_on_SS}
\end{eqnarray}
An examination of the terms in Eq.~(\ref{A_on_SS}) shows that the coefficient of $\bm{Y}$ is just the Jacobian, that is,
\begin{equation}
\left. A_I(\bm{y})\right|_{\rm SS} = \epsilon \sum^{2 \mathcal{D}}_{J=1} J_{I J}\,Y_{J} + \ldots,
\label{A_on_SS_part_uni}
\end{equation}
where the dots signify the terms in Eq.~(\ref{A_on_SS}) which do not involve $\bm{Y}$. This suggests that $\bm{Y}$ should be decomposed as follows:
\begin{equation}
Y_I = \sum^{2\mathcal{D}}_{K=2} w^{\{ K\} } V^{\{K\}}_I ,
\label{decomposition}
\end{equation}
with the $K=1$ term giving no contribution since $\sum^{2 \mathcal{D}}_{J=1} J_{I J}\,V^{\{ 1\} }_J = 0$. Then
\begin{equation}
\left. A_I(\bm{y})\right|_{\rm SS} = \epsilon \sum^{2 \mathcal{D}}_{K=2} \lambda^{\{ K\} }\,w^{\{ K\} } V^{\{ K\} }_I + \ldots.
\label{A_on_SS_part_uni_2}
\end{equation}

The condition that $\bm{A}(\bm{y})$ has no components in the fast directions $V^{\{ M\} }$, $M=2,\ldots,2\mathcal{D}$, can be written in the form $0 = \sum^{2\mathcal{D}}_{I=1} U^{\{M\} }_I A_I(\bm{y})$, $M=2,\ldots,2\mathcal{D}$. This shows why the form (\ref{A_on_SS_part_uni_2}) is useful: the $w^{(K)}$ are determined immediately by orthonormality, giving $0 = \epsilon \lambda^{\{M\} } w^{\{ M\} } + \ldots$. To make progress with the remaining terms, indicated by the dots, we need to break up the condition which determines the $w^{\{K\} }$:
\begin{equation}
0 = \sum^{2\mathcal{D}}_{I=1} U^{\{M\} }_I A_I(\bm{y}) = 
\sum^{\mathcal{D}}_{i=1} U^{\{M\} }_i A^{(1)}_{i}(\bm{y}) + 
\sum^{\mathcal{D}}_{i=1} U^{\{M\} }_{\mathcal{D}+i} A^{(2)}_{i}(\bm{y}),
\label{SS_condition}
\end{equation}
where $M=2,\ldots,2\mathcal{D}$. This gives the $w^{\{M\} }$, $M=2,\ldots, 2\mathcal{D}$, as
\begin{eqnarray}
w^{\{ M\} } &=& - \frac{1}{\lambda^{\{M\} }}\,\sum^{\mathcal{D}}_{i=1} U^{\{ M\} }_i \frac{z}{\beta_i}\,\left\{ \left( b^{(0)}_i \hat{b}^{(1)}_i - d^{(0)}_i \hat{d}^{(1)}_i \right) 
- c^{(0)}_{i} \hat{c}^{(1 1)}_i \kappa z - c^{(0)}_{i} \hat{c}^{(1 2)}_i \kappa \left( 1 - z \right)\right\} \nonumber \\
&-& \frac{1}{\lambda^{\{ M\} }}\,\sum^{\mathcal{D}}_{i=1} U^{\{ M\} }_{\mathcal{D}+i} \frac{(1 - z)}{\beta_i}\,\left\{ \left( b^{(0)}_i \hat{b}^{(2)}_i - d^{(0)}_i \hat{d}^{(2)}_i \right) 
- c^{(0)}_i \hat{c}^{(2 2)}_i \kappa \left( 1 - z \right) - c^{(0)}_i \hat{c}^{(2 1)}_i \kappa z \right\}.\nonumber \\
\label{w_first}
\end{eqnarray}
 
So, in summary, if the coordinates of the slow-subspace are chosen as
\begin{eqnarray}
y^{(1)}_i &=& z + \epsilon \sum^{2\mathcal{D}}_{K=2} w^{\{K\} } V^{\{K\} }_i, \nonumber \\
y^{(2)}_i &=& \left( 1 - z \right) + \epsilon \sum^{2\mathcal{D}}_{K=2} w^{\{K\} } V^{\{K\} }_{\mathcal{D}+i},
\label{summary_form}
\end{eqnarray}
then the $w^{\{K\} }$ are given by Eq.~(\ref{w_first}). 

\medskip

\section{Model reduction II.~Construction of the reduced model}\label{sec:reduction_SM}

\medskip

Our focus in the rest of the paper is then on the reduced form of the model that describes the second stage of the dynamics starting at the point $z^{\rm CMIC}$, and reaching an axis, at which point one or other of the alleles fix. We can now begin to construct this reduced theory. 

\subsection{The neutral model}
\label{sec:neutral_SM}
We have already seen that applying the condition $y^{(1)}_i = z$, $y^{(2)}_i = (1-z)$ gives a line of fixed points in the neutral model, that is $\bm{A} = 0$; there is no deterministic dynamics along the CM. In addition, if we denote differentiation with respect to $\tau$ by a dot, then $\dot{y}^{(2)}_{i} =-\dot{y}^{(1)}_{i}=- \dot{z}$ on the CM. Application of the projection operator $P_{IJ}$ to the left-hand side of the original stochastic differential equation (\ref{SDE_full}) then gives
\begin{eqnarray}
\sum^{2\mathcal{D}}_{J=1} V^{\{ 1\} }_I U^{\{1\} }_J \frac{\mathrm{d}x_J}{\mathrm{d}\tau} 
&=& \sum^{\mathcal{D}}_{J=1} V^{\{1\} }_I U^{\{ 1\} }_J \kappa \frac{\mathrm{d}z}{\mathrm{d}\tau} - \sum^{2\mathcal{D}}_{J=\mathcal{D}+1} V^{\{ 1\} }_I U^{\{ 1\} }_J \kappa \frac{\mathrm{d}z}{\mathrm{d}\tau} \nonumber \\
&=& \sum^{\mathcal{D}}_{j=1} V^{\{ 1\} }_I u^{\{ 1\} }_j \kappa \frac{\mathrm{d}z}{\mathrm{d}\tau} = V^{\{ 1\} }_I \kappa \frac{\mathrm{d}z}{\mathrm{d}\tau},
\label{LHS_reduced_eqn}
\end{eqnarray}
where we have used the form for $U^{\{ 1\} }_J$ given in Eq.~({\ref{first_D_eigenvectors}), and also $\sum^{\mathcal{D}}_{j=1} u^{\{ 1\} }_j = 1$, from orthogonality with $\underline{v}^{\{ 1\} }$.

The projection operator can also be applied to the noise term on the right-hand side of Eq.~(\ref{SDE_full}) to give
\begin{equation}
\frac{1}{\sqrt{V}}\,\sum^{2\mathcal{D}}_{J=1} V^{\{ 1\} }_I U^{\{ 1\} }_J \eta_{J}(\tau) = \frac{V^{\{ 1\} }_{I}}{\sqrt{V}}\,\sum^{\mathcal{D}}_{j=1} u^{\{ 1\} }_j \left[ \left( 1 - z \right) \eta^{(1)}_{j}(\tau) - z \eta^{(2)}_{j}(\tau) \right].
\label{P_on_noise}
\end{equation}
So the reduced stochastic differential equation in the neutral case may be written as
\begin{equation}
\frac{\mathrm{d}z}{\mathrm{d}\tau} = \frac{1}{\sqrt{V}} \zeta(\tau),
\label{SDE_neutral_reduced}
\end{equation}
where
\begin{equation}
\zeta(\tau) = \kappa^{-1}\sum^{\mathcal{D}}_{j=1} u^{\{ 1\} }_j \left[ \left( 1 - z \right) \eta^{(1)}_{j}(\tau) - z \eta^{(2)}_{j}(\tau) \right].
\label{reduced_SDE_neutral}
\end{equation}

It should be noted that since the noise depends on $z$, the direction of the dominant noise component changes along the CM. From the properties of $\eta_I$, we see that the effective noise $\zeta$ is Gaussian with zero mean and with correlator 
\begin{eqnarray}
& & \left\langle \zeta(\tau) \zeta(\tau') \right\rangle = \kappa^{-2}\sum^{\mathcal{D}}_{i,j=1} u^{\{ 1\} }_{i} u^{\{ 1\} }_{j} \left[ (1-z)^2 B^{(11)}_{ij} \right. 
\nonumber \\
&-& \left. z(1-z) B^{(12)}_{ij} - z(1-z) B^{(21)}_{ij} + z^2 B^{(22)}_{ij} \right] \delta\left( \tau - \tau' \right), \nonumber \\
\label{zeta_correlator}
\end{eqnarray}
with the $B_{IJ}$ being evaluated on the CM. From Eqs.~\eqref{part_of_B}--\eqref{rest_of_B}, with $x^{(1)}_i = \kappa z$ and $x^{(2)}_i = \kappa(1-z)$, one finds that 
\begin{eqnarray}
B^{(11)}_{ii}(z) &=& \frac{2\kappa z}{\beta^2_i}\,\left[ b^{(0)}_i + \sum_{j \neq i} \mu_{i j} \right]\,, \nonumber \\
B^{(22)}_{ii}(z) &=& \frac{2\kappa (1 - z)}{\beta^2_i}\,\left[ b^{(0)}_i + \sum_{j \neq i} \mu_{i j} \right]\,, \nonumber \\
B^{(11)}_{ij}(z) &=& - \frac{\kappa z}{\beta_i \beta_j}\,\left[ \mu_{i j} + \mu_{j i} \right] \ \ \left( i \neq j \right)\,, \nonumber \\
B^{(22)}_{ij}(z) &=& - \frac{\kappa (1 - z)}{\beta_i \beta_j}\,\left[ \mu_{i j} + \mu_{j i} \right] \ \ \left( i \neq j \right)\,,
\label{reduced_B}
\end{eqnarray}
with $B^{(12)}_{ij}=0$ and $B^{(21)}_{ij}=0$. A calculation of the term in square brackets in Eq.~(\ref{zeta_correlator}), allows us to arrive at the following form for the stochastic differential equation describing the neutral dynamics after the fast-mode elimination:
\begin{equation}
\frac{\mathrm{d}z}{\mathrm{d}\tau} = \bar{A}(z) + \frac{1}{\sqrt{V}} \zeta(\tau),
\label{SM_SDE_reduced}
\end{equation}
where $\bar{A}(z)=0$ and where $\zeta(\tau)$ is a Gaussian noise with zero 
mean and correlator
\begin{equation}
\left\langle \zeta(\tau) \zeta(\tau') \right\rangle = \bar{B}(z) \delta\left( \tau - \tau' \right)\,,
\label{SM_correlator_reduced}
\end{equation}
and where
\begin{eqnarray}
\bar{B}(z) &=& 2\kappa^{-1} z\left( 1 - z \right) \left\{ \sum^{\mathcal{D}}_{i=1} \frac{\left[ u^{\{ 1\} }_i \right]^{2}}{\beta^2_i} b^{(0)}_i - \sum^{\mathcal{D}}_{i,j = 1} \frac{  u^{\{ 1\} }_i u^{\{ 1\} }_j}{\beta_j} H_{i j} \right\} \nonumber \\
&=&  2\kappa^{-1} z\left( 1 - z \right) \sum^{\mathcal{D}}_{i=1} \frac{\left[ u^{\{ 1\} }_i \right]^{2}}{\beta^2_i} b^{(0)}_i,
\label{form_of_Bbar}
\end{eqnarray}
since $\sum_{i} u^{\{ 1\} }_i H_{i j}=0$.

Figure~\ref{fig:phase_neutral} of the main text shows a phase diagram of the dynamics of a neutral system with $\mathcal{D}=2$ islands in terms of the population of individuals of both alleles on one of the islands, while Fig.~\ref{fig:phase_neutral_SM} does it in terms of the population of individuals of one of the alleles in both islands. From these, we can observe the almost deterministic collapse of the stochastic system towards the CM given by $x_i^{(\alpha)} + x_i^{(\beta)} = \kappa$, with the values of a given $x_i^{(\alpha)}$ being independent of $i$. After that, the dynamics are only stochastic, reflecting the fact that $\bar{A}(z)=0$ in the neutral case.

Although the reduced neutral system given by Eqs.~(\ref{SM_SDE_reduced}) and (\ref{SM_correlator_reduced}), with $\bar{A}(z)=0$ and $\bar{B}(z)$ given by Eq.~(\ref{form_of_Bbar}), is of interest, the inclusion of selection gives a far richer structure. Since selection effects are weak, these can be included as perturbative corrections to the neutral theory just developed.

\subsection{The model with selection}
\label{sec:selection_SM}
To go on to analyse the non-neutral case we write the birth, death and competition parameters as in Eq.~(\ref{parameters_non_neut}) of the main text. We will keep order $\epsilon$ terms in $A_{I}(\bm{y})$, but only order one terms in $B_{IJ}(\bm{y})$ when carrying out the reduction. The reason for this is that we will tentatively assume that $\epsilon$ and $V^{-1}$ are essentially of the same order. This corresponds to keeping terms of order $\epsilon/V$ and $1/V^2$ in Eq.~(\ref{FPE_full}), but neglecting terms of order $\epsilon^2/V, \epsilon/V^2$ and $1/V^3$. Therefore the calculation of the noise correlator in the neutral theory carried out above is sufficient, and so all that is left is to find $A_{I}(\bm{y})$ on the SS to first order in $\epsilon$. 

To do this, we substitute Eq.~(\ref{decomposition}) into Eq.~(\ref{A_on_SS_part_uni}) to find:
\begin{equation}
\left. A_I(\bm{y})\right|_{\rm SS} = \epsilon \sum^{2 \mathcal{D}}_{K=2} w^{\{K\} } \lambda^{\{K\} } V^{\{K\} }_I  + \ldots,
\label{A_on_SS_after_sub}
\end{equation}
where the $\ldots$ once again refer to the terms in Eq.~(\ref{A_on_SS}) which do not involve $Y_I$. However, when we operate on $\left. A_I(\bm{y})\right|_{\rm SS}$ with the projection operator $P_{J I} = V^{\{1\} }_J U^{\{1\} }_I$ we get zero for the contribution shown in Eq.~(\ref{A_on_SS_after_sub}), since $\sum^{2\mathcal{D}}_{I=1} U^{\{ 1\} }_I V^{\{K\} }_I = 0$ for $K \geq 2$. Therefore the terms involving $Y_I$ in Eq.~(\ref{A_on_SS}) give no contribution. This means that to determine $\bar{A}(z)$ we only need in effect to consider 
\begin{eqnarray}
\left. A^{(1)}_{i}(\bm{y})\right|_{\rm SS} &=& 
\frac{\epsilon}{\beta_i} z \left\{ \left( b^{(0)}_i \hat{b}^{(1)}_i - d^{(0)}_i \hat{d}^{(1)}_i \right) - c^{(0)}_i \hat{c}^{(1 1)}_i \kappa z - c^{(0)}_{i} \hat{c}^{(1 2)}_i \kappa \left( 1 - z \right)\right\} + \mathcal{O}\left( \epsilon^2 \right), \nonumber \\
\left. A^{(2)}_{i}(\bm{y})\right|_{\rm SS} &=& 
\frac{\epsilon}{\beta_i} \left( 1 - z \right)\,\left\{ \left( b^{(0)}_i \hat{b}^{(2)}_i - d^{(0)}_i \hat{d}^{(2)}_i \right) - c^{(0)}_i \hat{c}^{(2 2)}_i \kappa \left( 1 - z \right) - c^{(0)}_i \hat{c}^{(2 1)}_i \kappa z \right\} + \mathcal{O}\left( \epsilon^2 \right). \nonumber \\
\label{effective_A_on_SS}
\end{eqnarray}
If we now act with the projection operator $P_{J I} = V^{\{1\} }_J U^{\{1\} }_I$, and omit the $V^{\{1\} }_J$ (which is plus one for the first $\mathcal{D}$ entries and minus one for the last $\mathcal{D}$ entries), we find that 
\begin{eqnarray}
\bar{A}(z) &=& \epsilon z \sum^{\mathcal{D}}_{i=1} \frac{U^{\{1\} }_i}{\beta_i}\,\left[ \left( b^{(0)}_i \hat{b}^{(1)}_i - d^{(0)}_i \hat{d}^{(1)}_i \right) - c^{(0)}_i \hat{c}^{(1 1)}_i \kappa z 
- c^{(0)}_i \hat{c}^{(1 2)}_i \kappa \left( 1 - z \right)\right] \nonumber \\
&+& \epsilon \left( 1 - z \right) \sum^{\mathcal{D}}_{i=1} \frac{U^{\{1\} }_{\mathcal{D} + i}}{\beta_i}\,\left[\left( b^{(0)}_i \hat{b}^{(2)}_i - d^{(0)}_i \hat{d}^{(2)}_i \right)
- c^{(0)}_i \hat{c}^{(2 2 )}_i \kappa \left( 1 - z \right) - c^{(0)}_i \hat{c}^{(2 1)}_i \kappa z \right] + \mathcal{O}\left( \epsilon^2 \right),\nonumber \\
\label{Abar}
\end{eqnarray}
or using Eq.~(\ref{first_D_eigenvectors}) and rearranging slightly, this becomes
\begin{eqnarray}
\bar{A}(z) &=& \epsilon z \left( 1 - z \right) \sum^{\mathcal{D}}_{i=1} \frac{u^{\{1\} }_i}{\beta_i}\,\left\{ \left[ \left( b^{(0)}_i \hat{b}^{(1)}_i - d^{(0)}_i \hat{d}^{(1)}_i \right) - \left( b^{(0)}_i \hat{b}^{(2)}_i - d^{(0)}_i \hat{d}^{(2)}_i \right) \right] \right. \nonumber \\
&+& \left. \kappa c^{(0)}_i \left( \hat{c}^{(2 2)}_i - \hat{c}^{(1 2)}_i \right) - \kappa z c^{(0)}_i \left[ \hat{c}^{(1 1)}_i - \hat{c}^{(1 2)}_i - \hat{c}^{(2 1)}_i + \hat{c}^{(2 2)}_i \right] \right\} + \mathcal{O}\left( \epsilon^2 \right)\,.
\label{Abar_simpler}
\end{eqnarray}
This is given in the main text as Eqs.~(\ref{Abar_very_simple})--(\ref{a_two}).

Finally, we investigate how the model simplifies if we impose the condition that fixation occurs on the SS at $z=0$ and $z=1$, that is, that when $z=1$, $y^{(1)}_i=1$ and $y^{(2)}_i=0$, for all $i$ and that when $z=0$, $y^{(1)}_i=0$ and $y^{(2)}_i=1$, for all $i$. Using Eq.~(\ref{summary_form}), these conditions imply that
\begin{equation}
\left. \sum^{2\mathcal{D}}_{K=2} w^{\{K\} } V^{\{K\} }_I \right|_{z=0,1} = 0, \ \ \ I=1,\ldots,2\mathcal{D}.
\label{conds_bounds}
\end{equation}
Multiplying by $U^{\{M\} }_I$ (either at $z=0$ or $z=1$ as appropriate---recall that the eigenvectors depend on $z$), summing over $I$, and using orthogonality, gives
\begin{equation}
\left. w^{\{M\} } \right|_{z=0,1} = 0, \ \ \ M=2,\ldots,2\mathcal{D}.
\label{conds_on_w}
\end{equation}
From Eq.~(\ref{w_first}) these conditions imply that the following two quantities vanish:
\begin{eqnarray}
& & \sum^{\mathcal{D}}_{i=1} \left. U^{\{M\} }_{\mathcal{D}+i} \right|_{z=0} \frac{1}{\beta_i}\,\left\{ \left( b^{(0)}_i \hat{b}^{(2)}_i - d^{(0)}_i \hat{d}^{(2)}_i \right) - \kappa c^{(0)}_i \hat{c}^{(2 2)}_i \right\} \nonumber \\
& & \sum^{\mathcal{D}}_{i=1} \left. U^{\{M\} }_i \right|_{z=1} \frac{1}{\beta_i}\,\left\{ \left( b^{(0)}_i \hat{b}^{(1)}_i - d^{(0)}_i \hat{d}^{(1)}_i \right) - \kappa c^{(0)}_i \hat{c}^{(1 1)}_i \right\}\,. \nonumber \\
\label{conds_2}
\end{eqnarray}
Using Eq.~(\ref{eigenvectors}) we see that the conditions for $M \leq \mathcal{D}$ become trivial, whereas those for $M = m + \mathcal{D}$, $m=1,\ldots,\mathcal{D}$ may be written as
\begin{eqnarray}
\sum^{\mathcal{D}}_{i=1} \beta^{\{m\} }_{L,i} \frac{1}{\beta_i}\,\left\{ \left( b^{(0)}_i \hat{b}^{(2)}_i - d^{(0)}_i \hat{d}^{(2)}_i \right) - \kappa c^{(0)}_i \hat{c}^{(2 2)}_i \right\} &=& 0, \nonumber \\
\sum^{\mathcal{D}}_{i=1} \beta^{\{ m\} }_{L,i} \frac{1}{\beta_i}\,\left\{ \left( b^{(0)}_i \hat{b}^{(1)}_i - d^{(0)}_i \hat{d}^{(1)}_i \right) - \kappa c^{(0)}_i \hat{c}^{(1 1)}_i \right\} &=& 0. \nonumber \\
\label{conds_3}
\end{eqnarray}
Since the $\underline{\beta}^{\{m\} }_{L}$ are linearly independent, Eq.~(\ref{conds_end_points}) of the main text follows. Under these conditions the results given by Eqs.~(\ref{Abar_very_simple})--(\ref{a_two}) of the main text can be written in the form (\ref{Abar_like_Ref_Oned}) with the effective parameters given by Eq.~(\ref{effective_paras}).

\medskip

\section{Analysis of the reduced model}\label{sec:TQ_SM}

\medskip

To calculate the fixation probability and mean time to fixation, we revert to the formalism of Fokker-Planck equations. The one-dimensional It\={o} stochastic differential equation (\ref{SDE_reduced}) is equivalent to the Fokker-Planck equation~\cite{gardiner2009a,risken1989a}
\begin{equation}
\frac{\partial \bar{P}(z,t)}{\partial t} =
- \frac{1}{V}\,\frac{\partial }{\partial z} 
\left[ \bar{A}(z) \bar{P}(z,t) \right] 
+ \frac{1}{2V^2}\frac{\partial^2 }{\partial z^2} \left[ \bar{B}(z) \bar{P}(z,t) \right], 
\label{FPE_reduced}
\end{equation}
where $\bar{P}(z,t)$ is the probability distribution function of the reduced system. Rather than the forward equation (\ref{FPE_reduced}), it is its adjoint, the backward Fokker-Planck equation~\cite{gardiner2009a,risken1989a} 
\begin{equation}
\frac{\partial \bar{Q}(z,t)}{\partial t} =
\frac{\bar{A}(z)}{V}\,\frac{\partial \bar{Q}(z,t)}{\partial z} 
+ \frac{\bar{B}(z)}{2V^2}\frac{\partial^2 \bar{Q}(z,t)}{\partial z^2}, 
\label{BFPE_reduced}
\end{equation}
that is used in the calculation of fixation properties. 

From the general theory of backward Fokker-Planck equations~\cite{gardiner2009a,risken1989a} it follows that the probability of fixation of the first allele, which we denote by $Q(z_0)$, satisfies the ordinary differential equation
\begin{equation}
\frac{\bar{A}(z_0)}{V}\,\frac{\mathrm{d}Q(z_0)}{\mathrm{d}z_0} + \frac{\bar{B}(z_0)}{2V^2}\frac{\mathrm{d}^{2}Q(z_0)}{\mathrm{d}z^2_0} = 0,
\label{ODE_prob}
\end{equation}
with boundary conditions $Q(0)=0$ and $Q(1)=1$. The variable appearing in the equation is $z_0$, the initial value on the SS, since the backward equation has as its variable the initial value of the variable appearing in the Fokker Planck equation. In Eq.~(\ref{z_initial_condition}) this was referred to as $z^{\rm CMIC}$, but it will be denoted by $z_0$ here, since there should be no confusion with the $0$ label used earlier for neutral quantities. The boundary conditions can be understood as follows: if the system starts at $z=0$ there is no probability of fixation of allele $1$, whereas if it starts at $z=1$, allele $1$ is sure to fix.

The mean time to fixation (of either allele), which we denote by $T(z_0)$, satisfies the ordinary differential equation~\cite{gardiner2009a,risken1989a}
\begin{equation}
\frac{\bar{A}(z_0)}{V}\,\frac{\mathrm{d}T(z_0)}{\mathrm{d}z_0} + \frac{\bar{B}(z_0)}{2V^2}\frac{\mathrm{d}^{2}T(z_0)}{\mathrm{d}z^2_0} = -1,
\label{ODE_time}
\end{equation}
with boundary conditions $T(0)=0$ and $T(1)=0$. Here the boundary conditions can be understood by noting that if the system starts either $z=0$ or $z=1$, then the system immediately fixes to either allele $1$ or allele $2$.

In the neutral case ($\epsilon = 0$, which implies $\bar{A}=0$), it is found that~\cite{crow2009a}
\begin{eqnarray}
Q(z_0) &=& z_0, \label{Q_0} \\
T(z_{0}) &=& - V^2 b^{-1} \left[ (1-z_{0})\ln{(1 - z_{0})} + z_0 \ln{(z_0)} \right]\,.  
\label{T_0}
\end{eqnarray}
These analytical results are compared against simulations of the original $2\mathcal{D}$-dimensional microscopic system---obtained as the mean of a large number of realisations of the process---in Figs.~\ref{fig:TQ_2D} of the main text and Fig.~\ref{fig:TQ_4D} for the cases of $\mathcal{D}=2$ and $\mathcal{D}=4$ islands, respectively. We find that the agreement between theory and simulation is excellent.

%%%%%%%%%%%%%%%%%%%%%%%%%%%%%%%%%%%%%%%%%%%%%%%%%%%%%%%%%%%%%%%%%%%%%%%%%%
%%%%%%%%%%%%%%%%%%%%%%%%%%%%%%%%%%%%%%%%%%%%%%%%%%%%%%%%%%%%%%%%%%%%%%%%%%

\begin{figure}
\centering
\includegraphics[width=0.48\columnwidth]{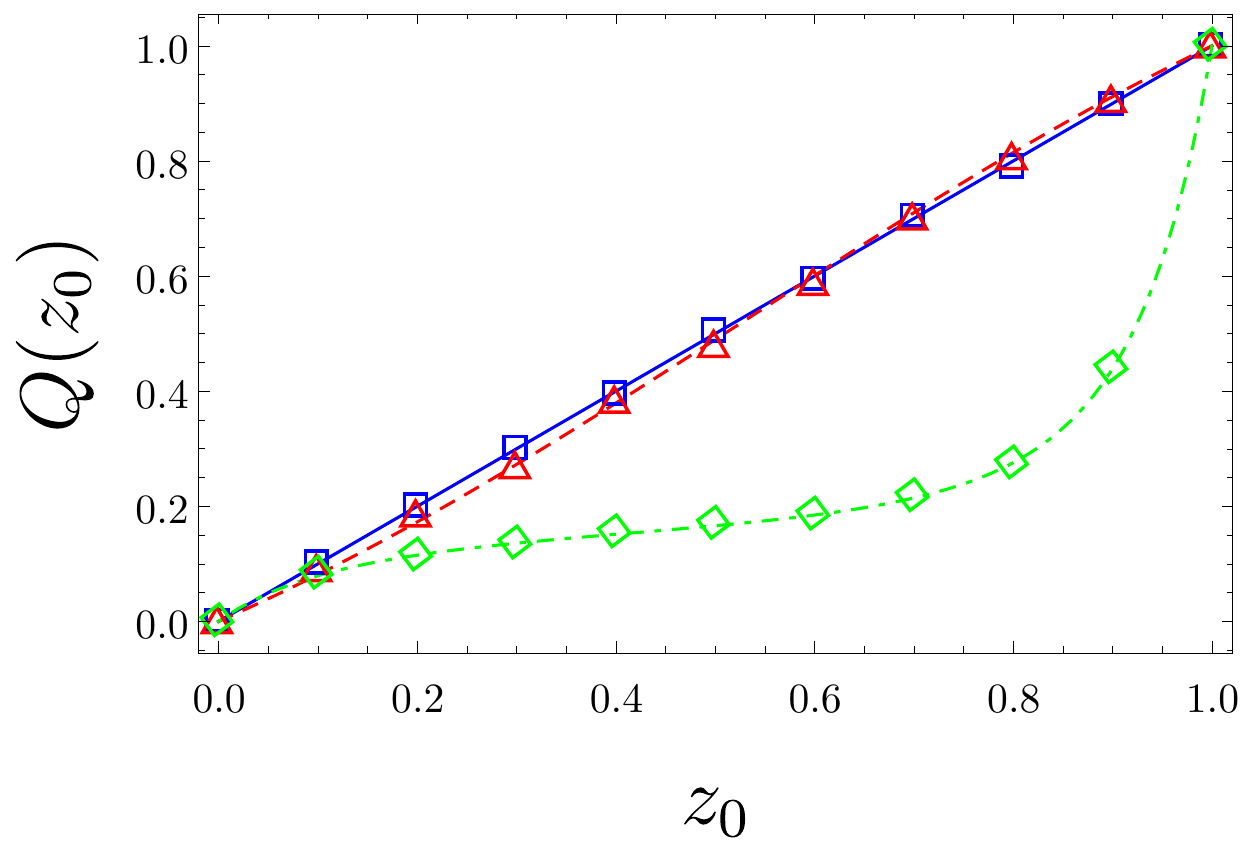}\hfill
\includegraphics[width=0.48\columnwidth]{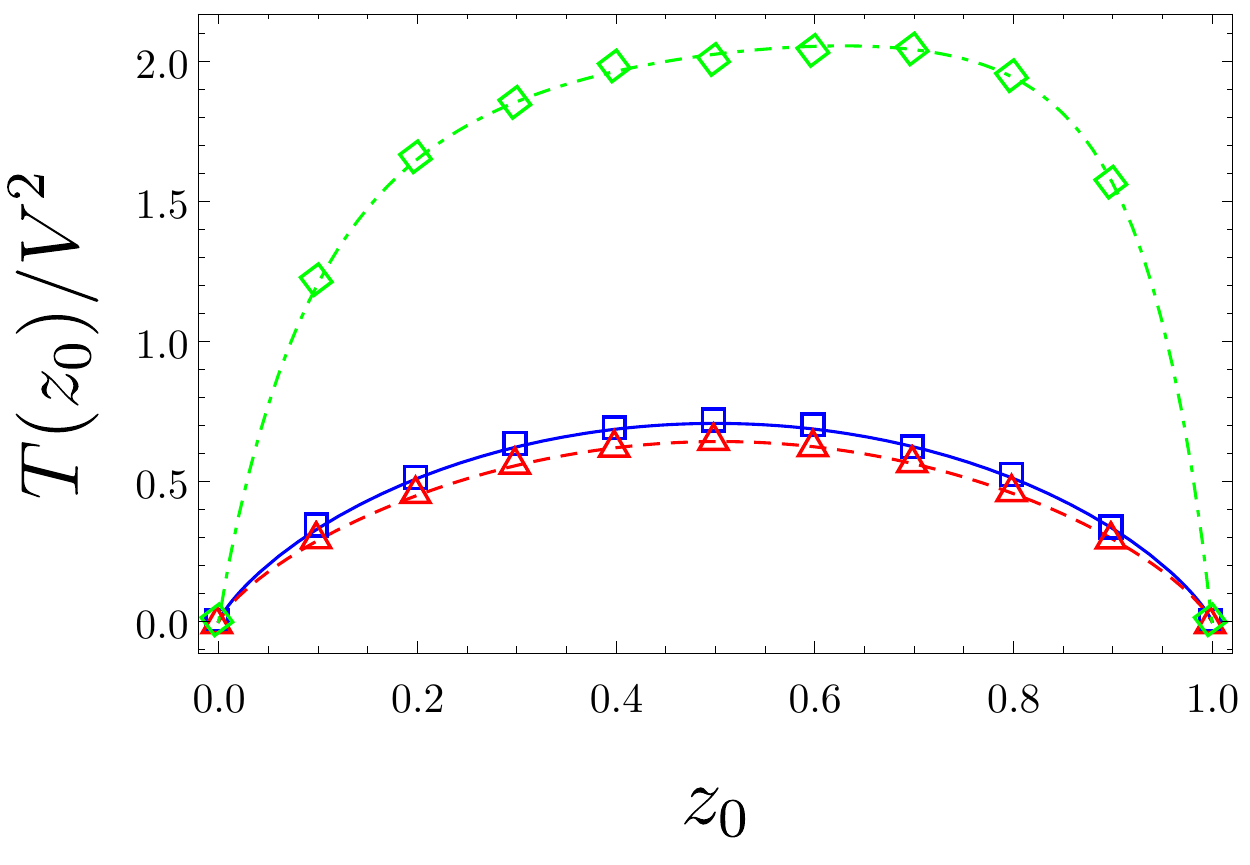}
\caption{Fixation probability of allele $1$ (top) and mean unconditional time to fixation (bottom) as a function of the projected initial condition $z_0$ for a system with $\mathcal{D}=4$, $V=150$, and $\kappa=1.5$. Blue (squares): neutral case; red (triangles, dashed): case with selection showing an unstable internal fixed point, with $\phi_{\rm eff}^{(1)}\approx -0.19$, $\phi_{\rm eff}^{(2)}\approx -0.17$, $\Gamma_{\rm eff}\approx -0.36$, and $z^*\approx 0.53$; green (diamonds, dot-dashed): case with selection showing a stable internal fixed point, with $\phi_{\rm eff}^{(1)}\approx 1.35$, $\phi_{\rm eff}^{(2)}\approx 1.9$, $\Gamma_{\rm eff}=3.25$, and $z^*\approx 0.42$. Symbols are obtained as the mean of 20000 stochastic simulations of the microscopic system, while the lines correspond to the theoretical predictions for the fixation probability and mean time to fixation, obtained from Eqs.~(\ref{Q_0}) and (\ref{T_0}) in the neutral case, and from Eq.~(\ref{Q_s}) and the analytical solution to Eq.~(\ref{ODE_time}) in the case with selection. The value of the selection parameter is $\epsilon=0.05$.}
\label{fig:TQ_4D}
\end{figure}

%%%%%%%%%%%%%%%%%%%%%%%%%%%%%%%%%%%%%%%%%%%%%%%%%%%%%%%%%%%%%%%%%%%%%%%%%%
%%%%%%%%%%%%%%%%%%%%%%%%%%%%%%%%%%%%%%%%%%%%%%%%%%%%%%%%%%%%%%%%%%%%%%%%%%

When selection is present, the calculation is less straightforward, but a relatively simple expression may be obtained for $Q(z_0)$. Following Ref.~\cite{constable2014ba}, if $\Gamma_{\rm eff} \neq 0$, we define
\begin{equation}
\ell(z_0) = \sqrt{\frac{V\epsilon}{2b|\Gamma_{\rm eff}| }}\left( \Gamma_{\rm eff} z_0 - \phi^{(1)}_{\rm eff}\right).
\label{ell_defn}
\end{equation}
Then it is found that
\begin{equation}
Q(z_0) = \frac{ 1 - \chi(z_0)}{1 - \chi(1)}; \ \ \ \ 
\chi(z_0) = \frac{f(l(z_0))}{f(l(0))}, 
\label{Q_s}
\end{equation}
where
\begin{eqnarray}
f(l(z_0)) = \mathrm{erfc}\left[ l(z_0) \right]\,, \quad \rm{if} \quad \Gamma_{\rm eff} < 0 \,,\nonumber \\
f(l(z_0)) = \mathrm{erfi}\left[ l(z_0) \right]\,, \quad \rm{if} \quad \Gamma_{\rm eff} > 0 \,. 
\label{eq_f}
\end{eqnarray}
Here erfc and erfi are respectively the complimentary and imaginary error functions~\cite{abramowitz1965a,erdelyi1953a}. If $\Gamma_{\rm eff}=0$, then $Q(z_0)$ still has the form $[1-\chi(z_0)][1 -\chi(1)]^{-1}$, but now $\chi(z_0) = \exp\{ - V\epsilon b^{-1}\phi^{(1)}_{\rm eff} z_0\}$. The calculation of $T(z_0)$ is more complex, and it is preferable to simply solve Eq.~(\ref{ODE_time}) numerically.

%%%%%%%%%%%%%%%%%%%%%%%%%%%%%%%%%%%%%%%%%%%%%%%%%%%%%%%%%%%%%%%%%%%%%%%%%%
%%%%%%%%%%%%%%%%%%%%%%%%%%%%%%%%%%%%%%%%%%%%%%%%%%%%%%%%%%%%%%%%%%%%%%%%%%

\begin{figure}
\centering
\includegraphics[width=0.48\columnwidth]{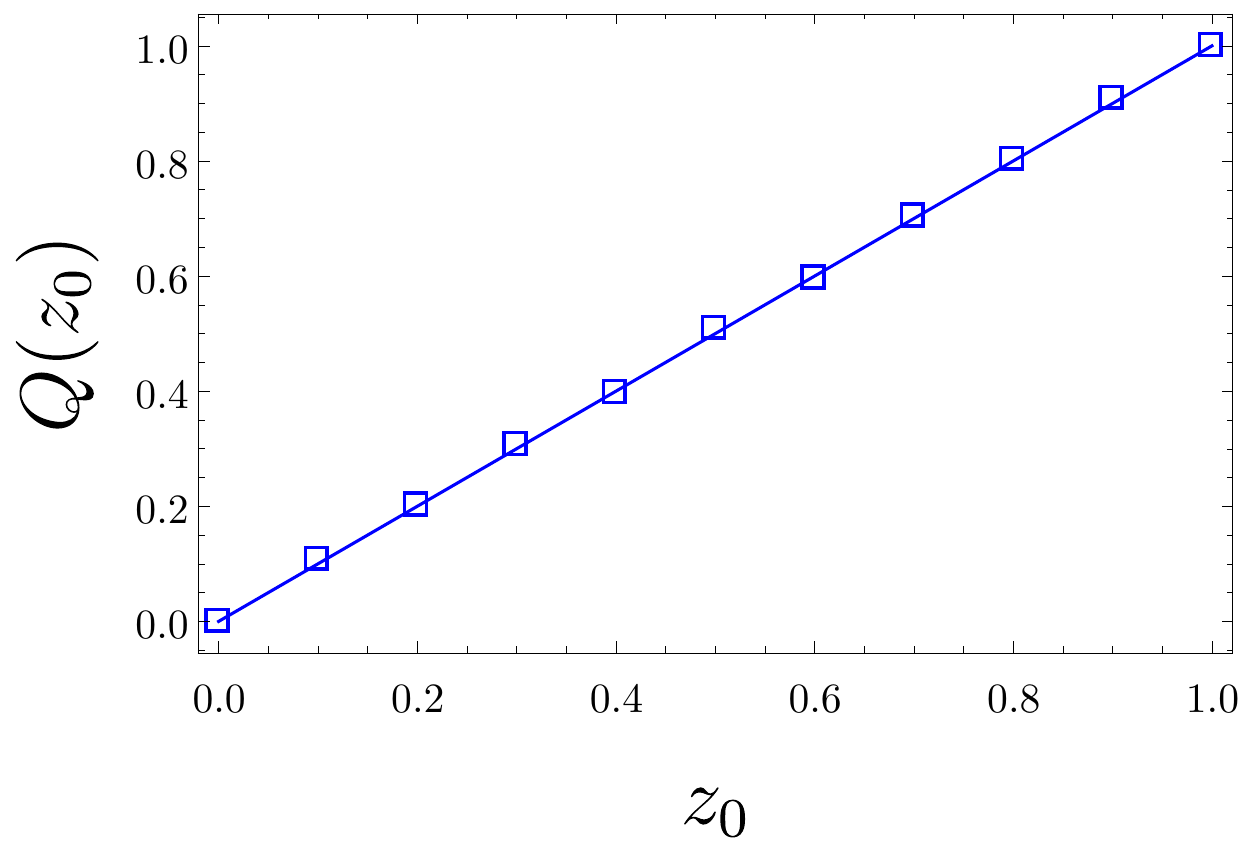}\hfill
\includegraphics[width=0.48\columnwidth]{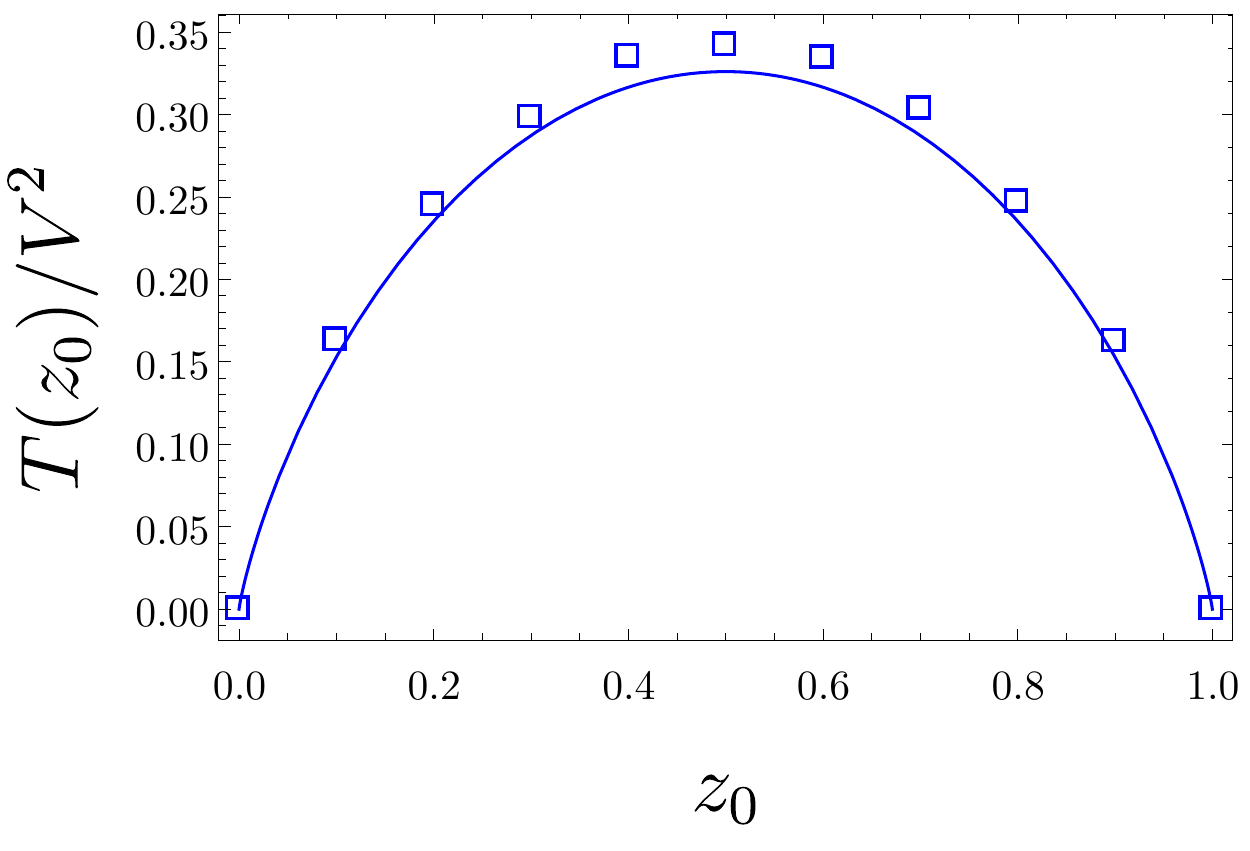}
\caption{Fixation probability of allele $1$ (top) and mean unconditional time to fixation (bottom) as a function of the projected initial condition $z_0$ for a neutral system with $\mathcal{D}=2$, $V=150$, and $\kappa=1.5$, when there is little separation between the magnitudes of the eigenvalues: $\lambda^{\{1\}}=0$, $\lambda^{\{2\}}\approx 0.013$, $\lambda^{\{3\}}\approx 0.07$, and $\lambda^{\{4\}}\approx 0.1$. Symbols: mean obtained from 10000 stochastic simulations of the microscopic system; lines: theoretical predictions for the fixation probability and mean time to fixation obtained from Eqs.~(\ref{Q_0}) and (\ref{T_0}), respectively.}
\label{fig:smallEVA}
\end{figure}

%%%%%%%%%%%%%%%%%%%%%%%%%%%%%%%%%%%%%%%%%%%%%%%%%%%%%%%%%%%%%%%%%%%%%%%%%%
%%%%%%%%%%%%%%%%%%%%%%%%%%%%%%%%%%%%%%%%%%%%%%%%%%%%%%%%%%%%%%%%%%%%%%%%%%

The results obtained from Eq.~(\ref{Q_s}) and the solution of Eq.~(\ref{ODE_time}) in the case with selection are again compared against simulations of the full system, and also shown in Fig.~\ref{fig:TQ_2D} of the main text and Fig.~\ref{fig:TQ_4D} for $\mathcal{D}=2$ and $\mathcal{D}=4$, respectively. In both cases, we compare the behaviour of the system with an unstable internal fixed point to that with a stable internal fixed point. Compared to the neutral case, an unstable fixed point results in a shorter time to fixation, and a stable fixed point in a longer time to fixation, as we had previously anticipated. Unlike the two-island scenario, where the signs of $\hat{c}^{(12)}_i$ and $\hat{c}^{(21)}_i$ were simply reversed to switch the stability of the fixed point, for the case with $\mathcal{D}=4$ shown in Fig.~\ref{fig:TQ_4D} their values have also been rescaled, due to the fact that simply switching them from positive to negative leads to fixation times more than an order of magnitude larger than in the neutral case.

Another aspect that is interesting to explore is the nature of the timescales involved in the collapse onto the SS (or the CM if there is no selection). We recall that the decay time of the various fast modes is proportional to the (magnitude of the real part of the) inverse of the eigenvalue of the Jacobian corresponding to that mode. In general the eigenvalues will depend on the parameters of the original model in a complicated way, and the only viable route to exploring their relative magnitudes is numerically. One question we can ask relates to the assumption of timescale separation on which the reduction method depends. Essentially the assumption is that there is a significant gap between the eigenvalues associated with the slow modes and those associated with the fast modes. This leads us to investigate parameter values for which there is little difference in the magnitude of eigenvalues of the system. That is, we ask: how does the reduced model perform in a case in which the timescale separation that justified the reduction in the first place is not so pronounced?

As mentioned in the main text, a disadvantage of the SLVC model is that it doubles the number of variables, as compared to the Moran model. It can nevertheless still be reduced to an effective one-variable model, just as in the case of the Moran model~\cite{constable2014aa,constable2014ba}. The structure of the fast modes is however more complex. It may be possible to find a set of parameters in which two sets of fast modes occur. For example, a faster set of $\mathcal{D}$ modes which involves a collapse from a system of $2\mathcal{D}$ variables to a $\mathcal{D}$ variable Moran type model, and then $\mathcal{D}-1$ slightly slower modes which would mirror the fast mode reduction of the Moran model~\cite{constable2014aa,constable2014ba}. Similarly, it might be possible to find another set of parameters where a faster set of $2\mathcal{D}-2$ modes reduce the full SLVC model to an effective one island SLVC model with two degrees of freedom, and then one slightly slower mode which would mirror the fast mode reduction of the well-mixed SLVC model~\cite{constable2015aa}. However, we expect that, for most combinations of parameter values, the different types of fast modes will be of a similar order and inextricably mixed. In this case no clear-cut Moran-type $\mathcal{D}$-island model or SLVC effective island mode will exist as an intermediate state.

One of the few analytic results concerning the magnitude of the eigenvalues is given in Sec.~\ref{sec:slow_modes_SM}, where we show that a subset of $\mathcal{D}$ of the eigenvalues of the system, which correspond to fast modes, are limited in magnitude by the minimum difference between birth and death rates---see Eq.~(\ref{eigenvalue_bound}), replacing $\kappa$ by $(b^{(0)}-d^{(0)})_{\rm{min}}/c^{(0)}_{\rm{min}}$. This suggests that taking a small value for $(b^{(0)}-d^{(0)})_{\rm{min}}$ could lead to eigenvalues with real parts whose magnitude is small. The other set of fast modes come from the part of the Jacobian directly proportional to the migration coefficients $\mu_{ij}$. With the above in mind, then, we carried out simulations of the microscopic model with small migration coefficients and $b_i^{(0)}\gtrsim d_i^{(0)}$. The results are shown in Fig.~\ref{fig:smallEVA} for a neutral system with $\mathcal{D}=2$ islands, with eigenvalues $\lambda^{\{1\}}=0$, $\lambda^{\{2\}}\approx 0.013$, $\lambda^{\{3\}}\approx 0.07$, and $\lambda^{\{4\}}\approx 0.1$. We see that, although the approximation is not as good as in the previous cases with more moderate parameter values, the agreement between theory and simulation is still very good.

%\bibliography{D_islands}

\begin{thebibliography}{22}%
\makeatletter
\providecommand \@ifxundefined [1]{%
 \@ifx{#1\undefined}
}%
\providecommand \@ifnum [1]{%
 \ifnum #1\expandafter \@firstoftwo
 \else \expandafter \@secondoftwo
 \fi
}%
\providecommand \@ifx [1]{%
 \ifx #1\expandafter \@firstoftwo
 \else \expandafter \@secondoftwo
 \fi
}%
\providecommand \natexlab [1]{#1}%
\providecommand \enquote  [1]{``#1''}%
\providecommand \bibnamefont  [1]{#1}%
\providecommand \bibfnamefont [1]{#1}%
\providecommand \citenamefont [1]{#1}%
\providecommand \href@noop [0]{\@secondoftwo}%
\providecommand \href [0]{\begingroup \@sanitize@url \@href}%
\providecommand \@href[1]{\@@startlink{#1}\@@href}%
\providecommand \@@href[1]{\endgroup#1\@@endlink}%
\providecommand \@sanitize@url [0]{\catcode `\\12\catcode `\$12\catcode
  `\&12\catcode `\#12\catcode `\^12\catcode `\_12\catcode `\%12\relax}%
\providecommand \@@startlink[1]{}%
\providecommand \@@endlink[0]{}%
\providecommand \url  [0]{\begingroup\@sanitize@url \@url }%
\providecommand \@url [1]{\endgroup\@href {#1}{\urlprefix }}%
\providecommand \urlprefix  [0]{URL }%
\providecommand \Eprint [0]{\href }%
\providecommand \doibase [0]{http://dx.doi.org/}%
\providecommand \selectlanguage [0]{\@gobble}%
\providecommand \bibinfo  [0]{\@secondoftwo}%
\providecommand \bibfield  [0]{\@secondoftwo}%
\providecommand \translation [1]{[#1]}%
\providecommand \BibitemOpen [0]{}%
\providecommand \bibitemStop [0]{}%
\providecommand \bibitemNoStop [0]{.\EOS\space}%
\providecommand \EOS [0]{\spacefactor3000\relax}%
\providecommand \BibitemShut  [1]{\csname bibitem#1\endcsname}%
\let\auto@bib@innerbib\@empty
%</preamble>
\bibitem [{\citenamefont {de~Viadar}\ and\ \citenamefont
  {Barton}(2011)}]{deViadar2011}%
  \BibitemOpen
  \bibfield  {author} {\bibinfo {author} {\bibfnamefont {H.~P.}\ \bibnamefont
  {de~Viadar}}\ and\ \bibinfo {author} {\bibfnamefont {N.~H.}\ \bibnamefont
  {Barton}},\ }\href@noop {} {\bibfield  {journal} {\bibinfo  {journal} {Trends
  Ecol. Evol.}\ }\textbf {\bibinfo {volume} {26}},\ \bibinfo {pages} {424}
  (\bibinfo {year} {2011})}\BibitemShut {NoStop}%
\bibitem [{\citenamefont {Black}\ and\ \citenamefont
  {McKane}(2012)}]{black2012}%
  \BibitemOpen
  \bibfield  {author} {\bibinfo {author} {\bibfnamefont {A.~J.}\ \bibnamefont
  {Black}}\ and\ \bibinfo {author} {\bibfnamefont {A.~J.}\ \bibnamefont
  {McKane}},\ }\href@noop {} {\bibfield  {journal} {\bibinfo  {journal} {Trends
  Ecol. Evol.}\ }\textbf {\bibinfo {volume} {27}},\ \bibinfo {pages} {337}
  (\bibinfo {year} {2012})}\BibitemShut {NoStop}%
\bibitem [{\citenamefont {Ewens}(1969)}]{ewens1969}%
  \BibitemOpen
  \bibfield  {author} {\bibinfo {author} {\bibfnamefont {W.~J.}\ \bibnamefont
  {Ewens}},\ }\href@noop {} {\emph {\bibinfo {title} {Population Genetics}}}\
  (\bibinfo  {publisher} {Methuen},\ \bibinfo {address} {London},\ \bibinfo
  {year} {1969})\BibitemShut {NoStop}%
\bibitem [{\citenamefont {Ewens}(2004)}]{ewens2004}%
  \BibitemOpen
  \bibfield  {author} {\bibinfo {author} {\bibfnamefont {W.~J.}\ \bibnamefont
  {Ewens}},\ }\href@noop {} {\emph {\bibinfo {title} {Mathematical Population
  Genetics: I. Theoretical Introduction}}}\ (\bibinfo  {publisher}
  {Springer-Verlag},\ \bibinfo {address} {Berlin},\ \bibinfo {year} {2004})\
  \bibinfo {note} {{S}econd edition}\BibitemShut {NoStop}%
\bibitem [{\citenamefont {Gardiner}(2009)}]{gardiner2009}%
  \BibitemOpen
  \bibfield  {author} {\bibinfo {author} {\bibfnamefont {C.~W.}\ \bibnamefont
  {Gardiner}},\ }\href@noop {} {\emph {\bibinfo {title} {{H}andbook of
  {S}tochastic {M}ethods}}},\ \bibinfo {edition} {{F}ourth}\ ed.\ (\bibinfo
  {publisher} {Springer},\ \bibinfo {address} {Berlin},\ \bibinfo {year}
  {2009})\BibitemShut {NoStop}%
\bibitem [{\citenamefont {Risken}(1989)}]{risken1989}%
  \BibitemOpen
  \bibfield  {author} {\bibinfo {author} {\bibfnamefont {H.}~\bibnamefont
  {Risken}},\ }\href@noop {} {\emph {\bibinfo {title} {The Fokker-Planck
  Equation - Methods of Solution and Applications}}},\ \bibinfo {edition}
  {{S}econd}\ ed.\ (\bibinfo  {publisher} {Springer},\ \bibinfo {address}
  {Berlin},\ \bibinfo {year} {1989})\BibitemShut {NoStop}%
\bibitem [{\citenamefont {Crow}\ and\ \citenamefont {Kimura}(2009)}]{crow2009}%
  \BibitemOpen
  \bibfield  {author} {\bibinfo {author} {\bibfnamefont {J.~F.}\ \bibnamefont
  {Crow}}\ and\ \bibinfo {author} {\bibfnamefont {M.}~\bibnamefont {Kimura}},\
  }\href@noop {} {\emph {\bibinfo {title} {{A}n {I}ntroduction to {P}opulation
  {G}enetics {T}heory}}}\ (\bibinfo  {publisher} {The Blackburn Press},\
  \bibinfo {address} {Caldwell, New Jersey, USA},\ \bibinfo {year}
  {2009})\BibitemShut {NoStop}%
\bibitem [{\citenamefont {Haken}(1983)}]{Haken1983}%
  \BibitemOpen
  \bibfield  {author} {\bibinfo {author} {\bibfnamefont {H.}~\bibnamefont
  {Haken}},\ }\href@noop {} {\emph {\bibinfo {title} {Synergetics}}}\ (\bibinfo
   {publisher} {Springer},\ \bibinfo {address} {Berlin},\ \bibinfo {year}
  {1983})\BibitemShut {NoStop}%
\bibitem [{\citenamefont {van Kampen}(1985)}]{vanKampen1985}%
  \BibitemOpen
  \bibfield  {author} {\bibinfo {author} {\bibfnamefont {N.~G.}\ \bibnamefont
  {van Kampen}},\ }\href@noop {} {\bibfield  {journal} {\bibinfo  {journal}
  {Phys. Reps.}\ }\textbf {\bibinfo {volume} {124}},\ \bibinfo {pages} {69}
  (\bibinfo {year} {1985})}\BibitemShut {NoStop}%
\bibitem [{\citenamefont {Wiggins}(2003)}]{wiggins2003}%
  \BibitemOpen
  \bibfield  {author} {\bibinfo {author} {\bibfnamefont {S.}~\bibnamefont
  {Wiggins}},\ }\href@noop {} {\emph {\bibinfo {title} {{I}ntroduction to
  {A}pplied {N}onlinear {D}ynamical {S}ystems and {C}haos}}}\ (\bibinfo
  {publisher} {Springer},\ \bibinfo {address} {New York},\ \bibinfo {year}
  {2003})\BibitemShut {NoStop}%
\bibitem [{\citenamefont {Fisher}(1930)}]{fisher1930}%
  \BibitemOpen
  \bibfield  {author} {\bibinfo {author} {\bibfnamefont {R.~A.}\ \bibnamefont
  {Fisher}},\ }\href@noop {} {\emph {\bibinfo {title} {The Genetical Theory of
  Natural Selection}}}\ (\bibinfo  {publisher} {Clarendon Press},\ \bibinfo
  {address} {Oxford},\ \bibinfo {year} {1930})\BibitemShut {NoStop}%
\bibitem [{\citenamefont {Wright}(1931)}]{wright1931}%
  \BibitemOpen
  \bibfield  {author} {\bibinfo {author} {\bibfnamefont {S.}~\bibnamefont
  {Wright}},\ }\href@noop {} {\bibfield  {journal} {\bibinfo  {journal}
  {Genetics}\ }\textbf {\bibinfo {volume} {16}},\ \bibinfo {pages} {97}
  (\bibinfo {year} {1931})}\BibitemShut {NoStop}%
\bibitem [{\citenamefont {Moran}(1958)}]{moran1958}%
  \BibitemOpen
  \bibfield  {author} {\bibinfo {author} {\bibfnamefont {P.~A.~P.}\
  \bibnamefont {Moran}},\ }\href@noop {} {\bibfield  {journal} {\bibinfo
  {journal} {Math. Proc. Camb. Philos. Soc.}\ }\textbf {\bibinfo {volume}
  {54}},\ \bibinfo {pages} {463} (\bibinfo {year} {1958})}\BibitemShut
  {NoStop}%
\bibitem [{\citenamefont {Pielou}(1977)}]{pielou1977}%
  \BibitemOpen
  \bibfield  {author} {\bibinfo {author} {\bibfnamefont {E.~C.}\ \bibnamefont
  {Pielou}},\ }\href@noop {} {\emph {\bibinfo {title} {Mathematical
  Ecology}}},\ \bibinfo {edition} {{S}econd}\ ed.\ (\bibinfo  {publisher}
  {Wiley},\ \bibinfo {address} {New York},\ \bibinfo {year} {1977})\BibitemShut
  {NoStop}%
\bibitem [{\citenamefont {Hanski}(1999)}]{hanski1999}%
  \BibitemOpen
  \bibfield  {author} {\bibinfo {author} {\bibfnamefont {I.}~\bibnamefont
  {Hanski}},\ }\href@noop {} {\emph {\bibinfo {title} {Metapopulation
  Ecology}}}\ (\bibinfo  {publisher} {Oxford University Press},\ \bibinfo
  {address} {Oxford},\ \bibinfo {year} {1999})\BibitemShut {NoStop}%
\bibitem [{\citenamefont {van Kampen}(2007)}]{vanKampen2007}%
  \BibitemOpen
  \bibfield  {author} {\bibinfo {author} {\bibfnamefont {N.~G.}\ \bibnamefont
  {van Kampen}},\ }\href@noop {} {\emph {\bibinfo {title} {{S}tochastic
  {P}rocesses in {P}hysics and {C}hemistry}}},\ \bibinfo {edition} {{T}hird}\
  ed.\ (\bibinfo  {publisher} {Elsevier Science},\ \bibinfo {address}
  {Amsterdam},\ \bibinfo {year} {2007})\BibitemShut {NoStop}%
\bibitem [{\citenamefont {Lombardo}\ \emph {et~al.}(2014)\citenamefont
  {Lombardo}, \citenamefont {Gambassi},\ and\ \citenamefont
  {Dall'Asta}}]{lombardo_2014}%
  \BibitemOpen
  \bibfield  {author} {\bibinfo {author} {\bibfnamefont {P.}~\bibnamefont
  {Lombardo}}, \bibinfo {author} {\bibfnamefont {A.}~\bibnamefont {Gambassi}},
  \ and\ \bibinfo {author} {\bibfnamefont {L.}~\bibnamefont {Dall'Asta}},\
  }\href@noop {} {\bibfield  {journal} {\bibinfo  {journal} {Phys. Rev. Lett.}\
  }\textbf {\bibinfo {volume} {112}},\ \bibinfo {pages} {148101} (\bibinfo
  {year} {2014})}\BibitemShut {NoStop}%
\bibitem [{\citenamefont {Lombardo}\ \emph {et~al.}(2015)\citenamefont
  {Lombardo}, \citenamefont {Gambassi},\ and\ \citenamefont
  {Dall'Asta}}]{lombardo_2015}%
  \BibitemOpen
  \bibfield  {author} {\bibinfo {author} {\bibfnamefont {P.}~\bibnamefont
  {Lombardo}}, \bibinfo {author} {\bibfnamefont {A.}~\bibnamefont {Gambassi}},
  \ and\ \bibinfo {author} {\bibfnamefont {L.}~\bibnamefont {Dall'Asta}},\
  }\href@noop {} {\bibfield  {journal} {\bibinfo  {journal} {Phys. Rev. E}\
  }\textbf {\bibinfo {volume} {91}},\ \bibinfo {pages} {032130} (\bibinfo
  {year} {2015})}\BibitemShut {NoStop}%
\bibitem [{\citenamefont {Constable}\ and\ \citenamefont
  {McKane}(2015{\natexlab{a}})}]{constable2015a}%
  \BibitemOpen
  \bibfield  {author} {\bibinfo {author} {\bibfnamefont {G.~W.~A.}\
  \bibnamefont {Constable}}\ and\ \bibinfo {author} {\bibfnamefont {A.~J.}\
  \bibnamefont {McKane}},\ }\href@noop {} {\bibfield  {journal} {\bibinfo
  {journal} {Phys. Rev. Lett.}\ }\textbf {\bibinfo {volume} {114}},\ \bibinfo
  {pages} {038101} (\bibinfo {year} {2015}{\natexlab{a}})}\BibitemShut
  {NoStop}%
\bibitem [{\citenamefont {Constable}\ and\ \citenamefont
  {McKane}(2014{\natexlab{a}})}]{constable2014a}%
  \BibitemOpen
  \bibfield  {author} {\bibinfo {author} {\bibfnamefont {G.~W.~A.}\
  \bibnamefont {Constable}}\ and\ \bibinfo {author} {\bibfnamefont {A.~J.}\
  \bibnamefont {McKane}},\ }\href@noop {} {\bibfield  {journal} {\bibinfo
  {journal} {Phys. Rev. E}\ }\textbf {\bibinfo {volume} {89}},\ \bibinfo
  {pages} {032141} (\bibinfo {year} {2014}{\natexlab{a}})}\BibitemShut
  {NoStop}%
\bibitem [{\citenamefont {Constable}\ and\ \citenamefont
  {McKane}(2014{\natexlab{b}})}]{constable2014b}%
  \BibitemOpen
  \bibfield  {author} {\bibinfo {author} {\bibfnamefont {G.~W.~A.}\
  \bibnamefont {Constable}}\ and\ \bibinfo {author} {\bibfnamefont {A.~J.}\
  \bibnamefont {McKane}},\ }\href@noop {} {\bibfield  {journal} {\bibinfo
  {journal} {J. Theor. Biol.}\ }\textbf {\bibinfo {volume} {358}},\ \bibinfo
  {pages} {149} (\bibinfo {year} {2014}{\natexlab{b}})}\BibitemShut {NoStop}%
\bibitem [{\citenamefont {Constable}\ and\ \citenamefont
  {McKane}(2015{\natexlab{b}})}]{constable2015c}%
  \BibitemOpen
  \bibfield  {author} {\bibinfo {author} {\bibfnamefont {G.~W.~A.}\
  \bibnamefont {Constable}}\ and\ \bibinfo {author} {\bibfnamefont {A.~J.}\
  \bibnamefont {McKane}},\ }\href@noop {} {\bibfield  {journal} {\bibinfo
  {journal} {Phys. Rev. E}\ }\textbf {\bibinfo {volume} {91}},\ \bibinfo
  {pages} {032711} (\bibinfo {year} {2015}{\natexlab{b}})}\BibitemShut
  {NoStop}%
\end{thebibliography}

\begin{thebibliography}{23}%
\makeatletter
\providecommand \@ifxundefined [1]{%
 \@ifx{#1\undefined}
}%
\providecommand \@ifnum [1]{%
 \ifnum #1\expandafter \@firstoftwo
 \else \expandafter \@secondoftwo
 \fi
}%
\providecommand \@ifx [1]{%
 \ifx #1\expandafter \@firstoftwo
 \else \expandafter \@secondoftwo
 \fi
}%
\providecommand \natexlab [1]{#1}%
\providecommand \enquote  [1]{``#1''}%
\providecommand \bibnamefont  [1]{#1}%
\providecommand \bibfnamefont [1]{#1}%
\providecommand \citenamefont [1]{#1}%
\providecommand \href@noop [0]{\@secondoftwo}%
\providecommand \href [0]{\begingroup \@sanitize@url \@href}%
\providecommand \@href[1]{\@@startlink{#1}\@@href}%
\providecommand \@@href[1]{\endgroup#1\@@endlink}%
\providecommand \@sanitize@url [0]{\catcode `\\12\catcode `\$12\catcode
  `\&12\catcode `\#12\catcode `\^12\catcode `\_12\catcode `\%12\relax}%
\providecommand \@@startlink[1]{}%
\providecommand \@@endlink[0]{}%
\providecommand \url  [0]{\begingroup\@sanitize@url \@url }%
\providecommand \@url [1]{\endgroup\@href {#1}{\urlprefix }}%
\providecommand \urlprefix  [0]{URL }%
\providecommand \Eprint [0]{\href }%
\providecommand \doibase [0]{http://dx.doi.org/}%
\providecommand \selectlanguage [0]{\@gobble}%
\providecommand \bibinfo  [0]{\@secondoftwo}%
\providecommand \bibfield  [0]{\@secondoftwo}%
\providecommand \translation [1]{[#1]}%
\providecommand \BibitemOpen [0]{}%
\providecommand \bibitemStop [0]{}%
\providecommand \bibitemNoStop [0]{.\EOS\space}%
\providecommand \EOS [0]{\spacefactor3000\relax}%
\providecommand \BibitemShut  [1]{\csname bibitem#1\endcsname}%
\let\auto@bib@innerbib\@empty
%</preamble>
\bibitem [{\citenamefont {Dobzhansky}(1937)}]{dobzhansky1937a}%
  \BibitemOpen
  \bibfield  {author} {\bibinfo {author} {\bibfnamefont {T.}~\bibnamefont
  {Dobzhansky}},\ }\href@noop {} {\emph {\bibinfo {title} {Genetics and the
  Origin of Species}}}\ (\bibinfo  {publisher} {Columbia Univ. Press},\
  \bibinfo {address} {New York},\ \bibinfo {year} {1937})\BibitemShut {NoStop}%
\bibitem [{\citenamefont {Mayr}(1942)}]{mayr1942a}%
  \BibitemOpen
  \bibfield  {author} {\bibinfo {author} {\bibfnamefont {E.}~\bibnamefont
  {Mayr}},\ }\href@noop {} {\emph {\bibinfo {title} {Systematics and the Origin
  of Species}}}\ (\bibinfo  {publisher} {Columbia Univ. Press},\ \bibinfo
  {address} {New York},\ \bibinfo {year} {1942})\BibitemShut {NoStop}%
\bibitem [{\citenamefont {Ewens}(2004)}]{ewens2004a}%
  \BibitemOpen
  \bibfield  {author} {\bibinfo {author} {\bibfnamefont {W.~J.}\ \bibnamefont
  {Ewens}},\ }\href@noop {} {\emph {\bibinfo {title} {Mathematical Population
  Genetics: I. Theoretical Introduction}}}\ (\bibinfo  {publisher}
  {Springer-Verlag},\ \bibinfo {address} {Berlin},\ \bibinfo {year} {2004})\
  \bibinfo {note} {{S}econd edition}\BibitemShut {NoStop}%
\bibitem [{\citenamefont {Roughgarden}(1979)}]{roughgarden_1979a}%
  \BibitemOpen
  \bibfield  {author} {\bibinfo {author} {\bibfnamefont {J.}~\bibnamefont
  {Roughgarden}},\ }\href@noop {} {\emph {\bibinfo {title} {{T}heory of
  {P}opulation {G}enetics and {E}volutionary {E}cology: {A}n {I}ntroduction}}}\
  (\bibinfo  {publisher} {Macmillan},\ \bibinfo {address} {New York},\ \bibinfo
  {year} {1979})\BibitemShut {NoStop}%
\bibitem [{\citenamefont {Wright}(1931)}]{wright1931a}%
  \BibitemOpen
  \bibfield  {author} {\bibinfo {author} {\bibfnamefont {S.}~\bibnamefont
  {Wright}},\ }\href@noop {} {\bibfield  {journal} {\bibinfo  {journal}
  {Genetics}\ }\textbf {\bibinfo {volume} {16}},\ \bibinfo {pages} {97}
  (\bibinfo {year} {1931})}\BibitemShut {NoStop}%
\bibitem [{\citenamefont {Kimura}\ and\ \citenamefont
  {Weiss}(1964)}]{kimuraSSMa}%
  \BibitemOpen
  \bibfield  {author} {\bibinfo {author} {\bibfnamefont {M.}~\bibnamefont
  {Kimura}}\ and\ \bibinfo {author} {\bibfnamefont {G.~H.}\ \bibnamefont
  {Weiss}},\ }\href@noop {} {\bibfield  {journal} {\bibinfo  {journal}
  {Genetics}\ }\textbf {\bibinfo {volume} {49}},\ \bibinfo {pages} {561}
  (\bibinfo {year} {1964})}\BibitemShut {NoStop}%
\bibitem [{\citenamefont {Maruyama}(1970)}]{maruyama1970a}%
  \BibitemOpen
  \bibfield  {author} {\bibinfo {author} {\bibfnamefont {T.}~\bibnamefont
  {Maruyama}},\ }\href@noop {} {\bibfield  {journal} {\bibinfo  {journal}
  {Genet. Res. Camb.}\ }\textbf {\bibinfo {volume} {15}},\ \bibinfo {pages}
  {221} (\bibinfo {year} {1970})}\BibitemShut {NoStop}%
\bibitem [{\citenamefont {Nagylaki}(1980)}]{nagylaki1980SMa}%
  \BibitemOpen
  \bibfield  {author} {\bibinfo {author} {\bibfnamefont {T.}~\bibnamefont
  {Nagylaki}},\ }\href@noop {} {\bibfield  {journal} {\bibinfo  {journal} {J.
  Math. Biol.}\ }\textbf {\bibinfo {volume} {9}},\ \bibinfo {pages} {101}
  (\bibinfo {year} {1980})}\BibitemShut {NoStop}%
\bibitem [{\citenamefont {Barton}(1993)}]{barton1993a}%
  \BibitemOpen
  \bibfield  {author} {\bibinfo {author} {\bibfnamefont {N.~H.}\ \bibnamefont
  {Barton}},\ }\href@noop {} {\bibfield  {journal} {\bibinfo  {journal} {Genet.
  Res.}\ }\textbf {\bibinfo {volume} {62}},\ \bibinfo {pages} {149} (\bibinfo
  {year} {1993})}\BibitemShut {NoStop}%
\bibitem [{\citenamefont {Whitlock}(2003)}]{whitlock2003a}%
  \BibitemOpen
  \bibfield  {author} {\bibinfo {author} {\bibfnamefont {M.~C.}\ \bibnamefont
  {Whitlock}},\ }\href@noop {} {\bibfield  {journal} {\bibinfo  {journal}
  {Genetics}\ }\textbf {\bibinfo {volume} {164}},\ \bibinfo {pages} {767}
  (\bibinfo {year} {2003})}\BibitemShut {NoStop}%
\bibitem [{\citenamefont {Rousset}(2004)}]{rousset2004a}%
  \BibitemOpen
  \bibfield  {author} {\bibinfo {author} {\bibfnamefont {F.}~\bibnamefont
  {Rousset}},\ }\href@noop {} {\emph {\bibinfo {title} {Genetic Structure and
  Selection in Subdivided Populations}}}\ (\bibinfo  {publisher} {Princeton
  University Press},\ \bibinfo {address} {Oxford},\ \bibinfo {year}
  {2004})\BibitemShut {NoStop}%
\bibitem [{\citenamefont {Constable}\ and\ \citenamefont
  {McKane}(2014{\natexlab{a}})}]{constable2014aa}%
  \BibitemOpen
  \bibfield  {author} {\bibinfo {author} {\bibfnamefont {G.~W.~A.}\
  \bibnamefont {Constable}}\ and\ \bibinfo {author} {\bibfnamefont {A.~J.}\
  \bibnamefont {McKane}},\ }\href@noop {} {\bibfield  {journal} {\bibinfo
  {journal} {Phys. Rev. E}\ }\textbf {\bibinfo {volume} {89}},\ \bibinfo
  {pages} {032141} (\bibinfo {year} {2014}{\natexlab{a}})}\BibitemShut
  {NoStop}%
\bibitem [{\citenamefont {Constable}\ and\ \citenamefont
  {McKane}(2014{\natexlab{b}})}]{constable2014ba}%
  \BibitemOpen
  \bibfield  {author} {\bibinfo {author} {\bibfnamefont {G.~W.~A.}\
  \bibnamefont {Constable}}\ and\ \bibinfo {author} {\bibfnamefont {A.~J.}\
  \bibnamefont {McKane}},\ }\href@noop {} {\bibfield  {journal} {\bibinfo
  {journal} {J. Theor. Biol.}\ }\textbf {\bibinfo {volume} {358}},\ \bibinfo
  {pages} {149} (\bibinfo {year} {2014}{\natexlab{b}})}\BibitemShut {NoStop}%
\bibitem [{\citenamefont {Constable}\ and\ \citenamefont
  {McKane}(2015)}]{constable2015aa}%
  \BibitemOpen
  \bibfield  {author} {\bibinfo {author} {\bibfnamefont {G.~W.~A.}\
  \bibnamefont {Constable}}\ and\ \bibinfo {author} {\bibfnamefont {A.~J.}\
  \bibnamefont {McKane}},\ }\href@noop {} {\bibfield  {journal} {\bibinfo
  {journal} {Phys. Rev. Lett.}\ }\textbf {\bibinfo {volume} {114}},\ \bibinfo
  {pages} {038101} (\bibinfo {year} {2015})}\BibitemShut {NoStop}%
\bibitem [{\citenamefont {van Kampen}(2007)}]{vanKampen2007a}%
  \BibitemOpen
  \bibfield  {author} {\bibinfo {author} {\bibfnamefont {N.~G.}\ \bibnamefont
  {van Kampen}},\ }\href@noop {} {\emph {\bibinfo {title} {{S}tochastic
  {P}rocesses in {P}hysics and {C}hemistry}}},\ \bibinfo {edition} {{T}hird}\
  ed.\ (\bibinfo  {publisher} {Elsevier Science},\ \bibinfo {address}
  {Amsterdam},\ \bibinfo {year} {2007})\BibitemShut {NoStop}%
\bibitem [{\citenamefont {Crow}\ and\ \citenamefont {Kimura}(2009)}]{crow2009a}%
  \BibitemOpen
  \bibfield  {author} {\bibinfo {author} {\bibfnamefont {J.~F.}\ \bibnamefont
  {Crow}}\ and\ \bibinfo {author} {\bibfnamefont {M.}~\bibnamefont {Kimura}},\
  }\href@noop {} {\emph {\bibinfo {title} {{A}n {I}ntroduction to {P}opulation
  {G}enetics {T}heory}}}\ (\bibinfo  {publisher} {The Blackburn Press},\
  \bibinfo {address} {Caldwell, New Jersey, USA},\ \bibinfo {year}
  {2009})\BibitemShut {NoStop}%
\bibitem [{\citenamefont {Gardiner}(2009)}]{gardiner2009a}%
  \BibitemOpen
  \bibfield  {author} {\bibinfo {author} {\bibfnamefont {C.~W.}\ \bibnamefont
  {Gardiner}},\ }\href@noop {} {\emph {\bibinfo {title} {{H}andbook of
  {S}tochastic {M}ethods}}},\ \bibinfo {edition} {{F}ourth}\ ed.\ (\bibinfo
  {publisher} {Springer},\ \bibinfo {address} {Berlin},\ \bibinfo {year}
  {2009})\BibitemShut {NoStop}%
\bibitem [{\citenamefont {Risken}(1989)}]{risken1989a}%
  \BibitemOpen
  \bibfield  {author} {\bibinfo {author} {\bibfnamefont {H.}~\bibnamefont
  {Risken}},\ }\href@noop {} {\emph {\bibinfo {title} {The Fokker-Planck
  Equation - Methods of Solution and Applications}}},\ \bibinfo {edition}
  {{S}econd}\ ed.\ (\bibinfo  {publisher} {Springer},\ \bibinfo {address}
  {Berlin},\ \bibinfo {year} {1989})\BibitemShut {NoStop}%
\bibitem [{\citenamefont {McKane}\ \emph {et~al.}(2014)\citenamefont {McKane},
  \citenamefont {Biancalani},\ and\ \citenamefont {Rogers}}]{mckane2014a}%
  \BibitemOpen
  \bibfield  {author} {\bibinfo {author} {\bibfnamefont {A.~J.}\ \bibnamefont
  {McKane}}, \bibinfo {author} {\bibfnamefont {T.}~\bibnamefont {Biancalani}},
  \ and\ \bibinfo {author} {\bibfnamefont {T.}~\bibnamefont {Rogers}},\
  }\href@noop {} {\bibfield  {journal} {\bibinfo  {journal} {Bull. Math.
  Biol.}\ }\textbf {\bibinfo {volume} {76}},\ \bibinfo {pages} {895} (\bibinfo
  {year} {2014})}\BibitemShut {NoStop}%
\bibitem [{\citenamefont {Gantmacher}(1959)}]{gantmacher1959a}%
  \BibitemOpen
  \bibfield  {author} {\bibinfo {author} {\bibfnamefont {F.~R.}\ \bibnamefont
  {Gantmacher}},\ }\href@noop {} {\emph {\bibinfo {title} {Applications of the
  Theory of Matrices}}}\ (\bibinfo  {publisher} {Interscience},\ \bibinfo
  {address} {New York},\ \bibinfo {year} {1959})\BibitemShut {NoStop}%
\bibitem [{\citenamefont {Cox}\ and\ \citenamefont {Miller}(1965)}]{cox1965a}%
  \BibitemOpen
  \bibfield  {author} {\bibinfo {author} {\bibfnamefont {D.~R.}\ \bibnamefont
  {Cox}}\ and\ \bibinfo {author} {\bibfnamefont {H.~D.}\ \bibnamefont
  {Miller}},\ }\href@noop {} {\emph {\bibinfo {title} {The Theory of Stochastic
  Processes}}}\ (\bibinfo  {publisher} {Chapman and Hall},\ \bibinfo {address}
  {London},\ \bibinfo {year} {1965})\BibitemShut {NoStop}%
\bibitem [{\citenamefont {Abramowitz}\ and\ \citenamefont
  {Stegun}(1965)}]{abramowitz1965a}%
  \BibitemOpen
  \bibinfo {editor} {\bibfnamefont {M.}~\bibnamefont {Abramowitz}}\ and\
  \bibinfo {editor} {\bibfnamefont {I.~A.}\ \bibnamefont {Stegun}},\ eds.,\
  \href@noop {} {\emph {\bibinfo {title} {{H}andbook of {M}athematical
  {F}unctions}}}\ (\bibinfo  {publisher} {Dover Publications},\ \bibinfo
  {address} {New York},\ \bibinfo {year} {1965})\BibitemShut {NoStop}%
\bibitem [{\citenamefont {Erd\'{e}lyi}(1953)}]{erdelyi1953a}%
  \BibitemOpen
  \bibinfo {editor} {\bibfnamefont {A.}~\bibnamefont {Erd\'{e}lyi}},\ ed.,\
  \href@noop {} {\emph {\bibinfo {title} {{H}igher {T}ranscendental
  {F}unctions: Vol II}}}\ (\bibinfo  {publisher} {McGraw-Hill},\ \bibinfo
  {address} {New York},\ \bibinfo {year} {1953})\BibitemShut {NoStop}%
\end{thebibliography}


%

%\end{document}

\end{document}